# Systematic study on probable projectile-target combinations for the synthesis of the $^{302}$120 superheavy nucleus

K. P. Santhosh and V. Safoora

*School of Pure and Applied Physics, Kannur University, Swami Anandatheertha Campus, Payyanur670327, Kerala, India*

Probable projectile-target combinations for the synthesis of superheavy element $^{302}$120 have been studied taking Coulomb and proximity potential as the interaction barrier. The probabilities of compound nucleus formation, $P_{CN}$ for the projectile-target combinations found in the cold reaction valley of $^{302}$120 are estimated. At energies near and above the Coulomb barrier, we have calculated the capture, fusion and evaporation residue cross sections for the reactions of all the probable projectile-target combinations so as to predict the most promising projectile-target combinations for the synthesis of SHE $^{302}$120 in heavy ion fusion reactions. The calculated fusion and evaporation cross section for the more asymmetric ("hotter") projectile-target combination is found to be higher than the less asymmetric ("colder") combination. It can be seen from the nature of quasi-fission barrier height, mass asymmetry, probability of compound nucleus formation, survival probability and excitation energy, the systems $^{44}$Ar + $^{258}$No, $^{46}$Ar + $^{256}$No, $^{48}$Ca + $^{254}$Fm, $^{50}$Ca + $^{252}$Fm, $^{54}$Ti + $^{248}$Cf, $^{58}$Cr + $^{244}$Cm in the deep region I of cold reaction valley, and the systems $^{62}$Fe + $^{240}$Pu, $^{64}$Fe + $^{238}$Pu, $^{68}$Ni + $^{234}$U, $^{70}$Ni + $^{232}$U, $^{72}$Ni + $^{230}$U, $^{74}$Zn + $^{228}$Th in the other cold valleys are identified as the better projectile-target combinations for the synthesis of $^{302}$120. Our prediction on the synthesis of $^{302}$120 superheavy nuclei using the combinations $^{54}$Cr+$^{248}$Cm, $^{58}$Fe+$^{244}$Pu, $^{64}$Ni+$^{238}$U and $^{50}$Ti+$^{249}$Cf are compared with available experimental data and other theoretical predictions.

email: drkpsanthosh@gmail.com

## I. INTRODUCTION

Heavy ion fusion reactions have been widely used to synthesize elements in the heavy and superheavy (SH) region and now nuclear reactions induced by heavy ions have become the principal tool in nuclear physics research. In order to form heavy nucleus, relatively heavier projectile must be fused with heavy target nuclei. This will lead to the formation of a highly excited completely fused system with a reduced probability of survival against fission. Up to now, considerable progress has been achieved in the experimental and theoretical investigations in the region of SHE [1-15]. Several successful [1] experiments have been done at different laboratories for the production of SHE with Z≤118 and an attempt to produce Z=120 have been reported [2]. To extend the periodic table, two different experimental approaches to synthesize SHE are used; the cold fusion reaction performed mainly at JINR-FLNR, Dubna and the hot fusion reaction performed mainly at GSI, Darmstadt, and at RIKEN, Japan. Using cold fusion approach, elements with Z=107-112 have been synthesized [4-6] and hot fusion reaction led to the discoveries of elements with Z=113-118[7-11]. Several theoretical calculations about the synthesis and decay of SHE have been performed within the fusion-by-diffusion (FBD) model[16-18], nuclear collectivization model [19-20], di-nuclear system(DNS) model [21-26] and Coulomb and Proximity Potential for Deformed Nuclei(CPPMDN)[27-29].

Nuclear theorists have predicted the existence of stability in the upper region of nuclear chart for the last four decade and are called the island of superheavy elements (SHE). The magic number for proton shell closure next to Z = 82 is predicted to be at Z =114,120 and 126 and that of neutron shell closure next to N = 126 is commonly predicted to be at N =184 [30-32]. The experimentalists have reached the shore of the island of stability around $Z$ = 120, 124, or 126 and N= 184 [33] through the progress in the accelerator technologies. The study of superheavy element Z=120 is of great interest, because it is useful in determining whether the magic proton shell should be at $Z$ = 114 or at higher proton numbers $Z$ = 120–126.

The short lifetimes and the low production cross sections (in the region of SHE, production cross section is in the order of pico barn) of the SH elements have posed difficulties to both experimentalists and theoreticians in studying the various properties of SH elements. In the calculation of the evaporation residue cross section, the reaction process leading to the synthesis of SHE can be divided into three steps. First, the projectile is captured by the target by overcoming the Coulomb barrier, which then evolves into the compound nucleus and, finally the compound nucleus loses excitation energy and cools down by the emission of particle and gamma rays and goes into the ground state. Therefore, the cross section for producing an evaporation residue, $\sigma_{ER}$ is the product of capture cross section $\sigma_{Capture}$, probability of compound nucleus formation $P_{CN}$ and the survival probability $W_{sur}$ of excited compound nucleus [34-36].

For light and medium-heavy projectile-target combination, where the fission barrier is high, each capture event leads to the formation of a compound nucleus, so $\sigma_{\mathrm{capture}} \approx \sigma_{\mathrm{fusion}} \approx \sigma_{\mathrm{ER}}$. But for heavy systems, especially for more symmetric combination, where the fission barrier is comparatively low and those leading to super heavy compound nuclei, the quasi fission process and the deep inelastic scattering, which lead to the two fragments in the exit channel, comes in and competes with the fusion process [35,37]. In this case, only a small part of the capture events are converted to fusion. Hence the complete fusion cross section ($\sigma_{fusion}/\sigma_{CN}$) is a part of the capture cross section $\sigma_{capture}$ and it is necessary to distinguish the capture and fusion processes [38]. The complete fusion cross section can be written as $\sigma_{fusion}= \sigma_{capture} \times P_{CN}$, where $P_{CN}$ is the probability that complete fusion will occur.

In 2008, experiments aimed at the synthesis of isotopes of element Z=120 have been done using $^{238}$U ($^{64}$Ni, xn) $^{302-x}$120 reaction and a cross section limit 90fb at E*=36.4 MeV was obtained [39]. Later in 2009, using the reaction $^{244}$Pu ($^{58}$Fe, xn) $^{302-x}$120, an attempt to produce Z=120 reaction have been performed, yielding an upper limit of 400fb [2]. A cross section limit of 560fb was established in the $^{54}$Cr+$^{248}$Cm reaction at SHIP in 2011[40, 41]. From the analysis of mass and total kinetic energy distributions, compound nucleus fission cross section were estimated experimentally by Kozulin et al., [42], and concluded that the combination $^{64}$Ni+$^{238}$U is less favorable to synthesis element Z = 120 compared to the combination $^{58}$Fe+$^{244}$Pu.

Several theoretical studies [43-56], are also being done for investigating the expected cross sections of yet unexplored reactions for the synthesis of new isotopes of element Z=120. Predicted empirical complete fusion cross sections for superheavy element Z = 120 for the system $^{252}$Cf ($^{50}$Ti, 4n) and $^{208}$Pb ($^{87}$Sr, n) by Loveland [43] is 30fb and 2fb respectively. Zagrebaev et al., [44] predicted that excitation functions for the production of the Z=120 element using $^{58}$Fe+$^{244}$Pu and $^{64}$Ni+$^{238}$U reactions are lower than those of the less symmetric $^{54}$Cr+$^{248}$Cm (see Fig.10 of Ref. [44]). The possibility to synthesis the element Z = 120 using $^{50}$Ti+$^{249-252}$Cf was evaluated by Liu et al.,[45] within the fusion-by-diffusion model and it was found that the reactions of $^{250,251}$Cf ($^{50}$Ti,3$n$) $^{297,298}$120 and $^{251,252}$Cf ($^{50}$Ti,4$n$) $^{297,298}$120 are relatively favorable with the maximum evaporation residue cross sections of 0.12pb, 0.09pb, 0.11pb, and 0.25 pb, respectively. Nasirov et al., [49] calculated and compared the fusion and evaporation residue cross sections of $^{50}$Ti+$^{249}$Cf and $^{54}$Cr+$^{248}$Cm systems and it was found that the evaporation residue excitation function for the more mass asymmetric reaction $^{50}$Ti+$^{249}$Cf is higher in comparison with less mass asymmetric $^{54}$Cr+$^{248}$Cm. The maximum values of the excitation function of the 3n channel of the evaporation residue formation of the above system are about 100fb and 55fb respectively. Siwek-Wilczynska et al., [52] predicted the cross sections for the reactions $^{249}$Cf ($^{50}$Ti, xn) $^{299-x}$120, for which the predicted cross section is only 6 fb. For producing Z = 120, the maximum ER cross section obtained by Bao et al., [55] is 0.12 pb by the 3n channel of $^{50}$Ti+$^{251}$Cf. The predicted maximum cross section varies greatly, depending on the models and methods used, the overall uncertainties in predicting the cross section of SHE is examined by Loveland [56].

The studies on heavy ion fusion cross section reveals that, the projectile-target combinations, the incident energy, mass asymmetry in the entrance channel and excitation energy are the key factors on which the fusions cross section strongly dependent. Hence it will be interesting and useful to study such dependencies for the synthesis of new superheavy elements (SHE). The theoretical predictions will be very useful for the experimentalist to choose the specific combinations with optimum energy and for the estimation of cross section. The selection of most promising projectile-target combination is crucial for an experimentalist for the synthesis of new element.

One of us (KPS) calculated the total fusion cross sections for the fusion of $^{12}$C, $^{16}$O, $^{28}$Si and $^{35}$Cl on $^{92}$Zr target [57]; $^{9}$Be on $^{27}$Al and $^{64}$Zn targets, $^{16}$O on $^{64}$Zn target [58]; $^{16}$O on $^{144-154}$Sm target [59] by taking scattering potential as the sum of Coulomb and proximity potential [60] and the computed values are compared with experimental data. By using the concept of cold reaction valley [61], probable projectile target combination for $^{286}$112 is also studied [62] and it was found that the systems $^{82}$Ge+$^{204}$Hg, $^{80}$Ge+$^{206}$Hg and $^{78}$Zn+$^{208}$Pb, $^{48}$Ca+$^{238}$U, $^{38}$S+$^{248}$Cm, $^{44}$Ar+$^{242}$Pu in the cold reaction valley are predicted to be the better optimal projectile target combinations for the synthesis of super heavy nuclei $^{286}$112. In the present work we have studied the cold reaction valley of $^{302}$120 SHE for identifying the better projectile-target combinations for the synthesis of $^{302}$120. The scattering potential energy curves for all probable combinations are studied. The probability of compound nucleus formation $P_{CN}$, capture cross sections $\sigma_{capture}$, fusion cross sections $\sigma_{fusion}$, survival probability $W_{sur}$ and evaporation residue cross section $\sigma_{ER}$ for the reactions of all the probable projectile-target combinations found in the cold valleys are systematically estimated. The details of scattering potential and the methodology used in the estimation of cross section are described in Section II. In Section III, results and discussion are given and the entire work is summarized in Section IV.

## II.THEORY

### A. THE POTENTIAL

The interaction barrier for the two colliding nuclei is given as:

$$V = \frac{Z_1 Z_2 e^2}{r} + V_P(z) + \frac{\hbar^2 \ell(\ell+1)}{2\mu r^2}, \qquad (1)$$

where $Z_1$ and $Z_2$ are the atomic numbers of projectile and target, $r$ is the distance between the centers of the projectile and target, $z$ is the distance between the near surfaces of the projectile and target, $\ell$ is the angular momentum, $\mu$ is the reduced mass of the target and projectile, and $V_P(z)$ is the proximity potential[63] given as:

$$V_P(z) = 4\pi\gamma b \frac{C_1 C_2}{C_1 + C_2} \phi(\frac{z}{b}), \quad (2)$$

with the nuclear surface tension coefficient,

$$\gamma = 0.9517[1 - 1.7826(N - Z)^2 / A^2], \quad (3)$$

$\phi$, the universal proximity potential is given as:

$$\phi(\xi) = -4.41\exp(-\xi/0.7176), \quad \text{for } \xi \geq 1.9475, \quad (4)$$

$$\phi(\xi) = -1.7817 + 0.9270\xi + 0.01696\xi^2 - 0.05148\xi^3, \quad \text{for } 0 \leq \xi \leq 1.9475, \quad (5)$$

$$\phi(\xi) = -1.7817 + 0.9270\xi + 0.0143\xi^2 - 0.09\xi^3, \quad \text{for } \xi \leq 0. \quad (6)$$

Here $\xi = z/b$, where the width (diffuseness) of nuclear surface $b \approx 1$ fm and Siissmann Central radii $C_i$ related to sharp radii $R_i$ as $C_i = R_i - \frac{b^2}{R_i}$.

For $R_i$, we use the semi empirical formula in terms of mass number $A_i$ as:

$$R_i = 1.28 A_i^{1/3} - 0.76 + 0.8 A_i^{-1/3}. \quad (7)$$

## B. CROSS SECTION

### 1. CAPTURE/TOTAL CROSS SECTION

To describe the fusion reactions at energies not too much above the barrier and at higher energies, the barrier penetration model developed by Wong [64] has been widely used, which obviously explains the experimental result properly.

The capture cross section at a given center of mass energy $E$ can be written as the sum of the cross section for each partial wave $\ell$,

$$\sigma_{capture} = \frac{\pi}{k^2} \sum_{\ell=0}^{\infty} (2\ell + 1) T(E, \ell), \quad (8)$$

where $T(E,\ell)$ denotes the penetration probability of $\ell^{th}$ partial wave, $k = \sqrt{\frac{2\mu E}{\hbar^2}}$ and $\mu$ is the reduced mass of the interacting system.

Following Thomas [65], Huizenga and Igo [66] and Rasmussen and Sugawara [67], Wong [60] approximated the various barriers for different partial waves by inverted harmonic oscillator potentials of height $E_\ell$ and frequency $\omega_\ell$. For energy $E$, the probability for the absorption of $\ell^{th}$ partial wave given by Hill-Wheeler formula [68] is

$$T(E, \ell) = \{1 + \exp[2\pi(E_\ell - E) / \hbar\omega_\ell]\}^{-1}. \quad (9)$$

In consequence, Wong arrived at the total cross section for the fusion of two nuclei and is given by

$$\sigma_{capture} = \frac{\pi}{k^2} \sum_{\ell} \frac{2\ell+1}{1+\exp[2\pi(E_\ell - E)/\hbar\omega_\ell]}. \tag{10}$$

Here $\hbar\omega_\ell$ is the curvature of the inverted parabola. Using some parameterizations in the region $\ell = 0$ and replacing the sum in Eq. (8) by an integral, Wong gave the reaction cross section as:

$$\sigma_{capture} = \frac{R_0^2 \hbar\omega_0}{2E} \ln\left\{1+\exp\left[\frac{2\pi(E-E_0)}{\hbar\omega_0}\right]\right\}, \tag{11}$$

where $R_0$ is the barrier radius and $E_0$ is the barrier height for $\ell = 0$.

The barrier radius for $\ell = 0$ is obtained from the condition

$$\left.\frac{dV(r)}{dr}\right|_{R_0} = 0. \tag{12}$$

The curvature $\hbar\omega_0$ is related to the potential given by

$$\hbar\omega_0 = \frac{\hbar}{\sqrt{\mu}} \sqrt{\left.\frac{d^2V(r)}{dr^2}\right|_{R_0}}. \tag{13}$$

For relatively large values of $E$, the above result reduces to the well-known formula:

$$\sigma_{capture} = \pi R_0^2 \left[1 - \frac{E_0}{E}\right]. \tag{14}$$

Lefort and collaborators have shown that not a critical angular momentum but a critical distance of approach may be the relevant quantity limiting complete fusion during a collision between two complex nuclei [69]. In order to substantiate the finding of a critical distance of approach, it is necessary to check the linear dependence of sigma on *1/E* in the region of high energy. The value of critical distance was found to be

$$R_C = r_C\left(A_1^{1/3} + A_2^{1/3}\right), \qquad r_C = 1.0 \pm 0.07 \text{ fm}. \tag{15}$$

Gutbrod, Winn, and Blann from their analysis of low energy data [70], obtain the fusion distance as

$$R_B = r_B\left(A_1^{1/3} + A_2^{1/3}\right), \qquad r_B = 1.4 \text{ fm}, \tag{16}$$

that is 40% larger than the value of $R_C$ and corresponds to the distance of the ions at the fusion barrier. In order to understand the difference between the two distances given by equations (15) and (16), Glas and Mosel [71] set cross section as

$$\sigma_{capture} = \frac{\pi}{k^2} \sum_{\ell=0}^{\infty} (2\ell+1) T_\ell P_\ell, \tag{17}$$

where $T_\ell$ is the penetration probability and $P_\ell = 1$ for $\ell \le \ell_c$; $P_\ell = 0$ for $\ell > \ell_c$.

Replacing the sum in Eq. (17) by an integration one obtains

$$\sigma_{capture} = \frac{R_B^2 \hbar\omega}{2} \frac{1}{E} \ln\left\{ \frac{1+\exp[2\pi(E-V(R_B)/\hbar\omega]}{1+\exp\left[2\pi\{E-V(R_B)-(R_C/R_B)^2[E-V(R_C)]\}/\hbar\omega\right]} \right\}, \tag{18}$$

where $V(R_B)$ and $V(R_C)$ are the barrier height corresponding to $R_B$ and $R_C$ respectively.

## 2. FUSION CROSS SECTION

We observe that the complete fusion cross section constitutes only a small fraction of the capture cross section, especially for the heavier targets and is always smaller or equal to the capture cross section. The fusion cross section is expressed as

$$\sigma_{fusion} = \frac{\pi}{k^2} \sum_{\ell=0}^{\infty} (2\ell+1) T(E,\ell) P_{CN}(E,\ell), \tag{19}$$

where $P_{CN}$ is the probability of compound nucleus formation which is described in next section.

### 2.1. PROBABILITY OF COMPOUND NUCLEUS FORMATION, $P_{CN}$

Different models and empirical formulas [43-44, 72-82] have been proposed for the calculation of most unclear part $P_{CN}$ and there is no satisfactory quantitative model for the fusion probability. Armbruster [73] has suggested

$$P_{CN}(E,\ell) = 0.5 \exp(-c(x_{eff} - x_{thr})) . \tag{20}$$

The studies [43-44, 74, 79] have found that there exists beam energy dependence for the fusion probability.

The energy dependent fusion probability is used by us to calculate $P_{CN}$ and is given by

$$P_{CN}(E,\ell) = \frac{\exp\{-c(x_{eff} - x_{thr})\}}{1+\exp\left\{\frac{E_B^* - E^*}{\Delta}\right\}}, \tag{21}$$

where $E^*$ is the excitation energy of the compound nucleus is, $E_B^*$ denotes the excitation energy of the CN when the center-of-mass beam energy is equal to the Coulomb and proximity barrier, $\Delta$ is an adjustable parameter ($\Delta = 4MeV$) and, and $x_{eff}$ is the effective fissility defined as,

$$x_{eff} = \left[ \frac{(Z^2/A)}{(Z^2/A)_{crit}} \right] \left(1 - \alpha + \alpha f(k)\right), \tag{22}$$

with $(Z^2/A)_{crit}$, $f(k)$ and $k$ is given by

$$(Z^2/A)_{crit} = 50.883 \left[ 1 - 1.7286 \left( \frac{(N-Z)}{A} \right)^2 \right], \tag{23}$$

$$f(k) = \frac{4}{K^2 + K + \frac{1}{k} + \frac{1}{k^2}}, \tag{24}$$

$$k = (A_1/A_2)^{1/3}, \tag{25}$$

where $Z$ , $N$ and $A$ represent the atomic number, neutron number and mass number respectively. $A_1$ and $A_2$ are mass number of projectile and target respectively. $x_{thr}$ , $c$ are adjustable parameters and $\alpha$ *=1/3*. The best fit to the cold fusion reaction, the values of *c* and $x_{eff}$ are *136.5* and *0.79* respectively. For hot fusion reaction, the best fit for $x_{eff} \leq 0.8$ is $c = 104$ and $x_{thr} = 0.69$; while $x_{eff} \geq 0.8$, the values are $c = 82$ and $x_{thr} = 0.69$. These constants are suggested by Loveland [43]. This form of energy dependence of fusion probability is similar to the one proposed by Zargrebeav [44].

## 3. EVAPORATION RESIDUE CROSS SECTION

The cross section of SH element production in a heavy ion fusion reaction with subsequent emission of *x* neutrons is given by

$$\sigma_{ER}^{xn} = \frac{\pi}{k^2} \sum_{\ell=0}^{\infty} (2\ell+1) T(E,\ell) P_{CN}(E,\ell) W_{sur}^{xn}(E^*,\ell) . \tag{26}$$

$W_{sur}$ is the probability for the compound nucleus to decay to the ground state of the final residual nucleus via evaporation of light particles and gamma ray for avoiding fission process and is described in next section.

### 3.1. SURVIVAL PROBABILITY, $W_{sur}$

The survival probability is the probability that the fused system emit several neutrons followed by observing a sequence of alpha decay from the residue. The survival probability under the evaporation of *x* neutrons is

$$W_{sur} = P_{xn}(E_{CN}^*) \prod_{i=1}^{i_{\max}=x} \left( \frac{\Gamma_n}{\Gamma_n + \Gamma_f} \right)_{i,E^*} , \tag{27}$$

where the index 'i' is equal to the number of emitted neutrons, $P_{xn}$ is the probability of emitting exactly *x* neutrons [83], $E^*$ is the excitation energy of the compound nucleus, $\Gamma_n$ and $\Gamma_f$ represent the decay width of neutron evaporation and fission respectively. To calculate $\Gamma_n / \Gamma_f$ , Vandenbosch and Huizenga [84] have suggested a classical formalism:

$$\frac{\Gamma_n}{\Gamma_f} = \frac{4A^{2/3} a_f (E^* - B_n)}{K_0 a_n [2a_f^{1/2} (E^* - B_f)^{1/2} - 1]} \exp[2a_n^{1/2} (E^* - B_n)^{1/2} - 2a_f^{1/2} (E^* - B_f)^{1/2}], \tag{28}$$

where A is the mass number of the nucleus considered, E* is the excitation energy, $B_n$ neutron separation energy. The constant $K_0$ is taken as 10MeV. $a_n = A/10$ and $a_f = 1.1 a_n$ , are the level density parameters of the daughter nucleus and the fissioning nucleus at the ground state and saddle configurations respectively. $B_f$ is the fission barrier and this height is a decisive quantity in the competition between processes of neutron evaporation and fission of the compound nucleus in the process of its cooling to form a residual nucleus in its ground state.

## III. RESULTS AND DISCUSSION

The driving potential, which is the difference between the interaction potential (Coulomb and proximity potential) and Q value of the reaction for all the projectile target combinations of SHE $^{302}$120 is calculated. For a fixed pair of ($A_P$, $A_T$), a pair of ($Z_P$, $Z_T$) is singled out for which the driving potential is minimum, and is plotted as a function of projectile mass number $A_P$. This plot is usually referred to as the cold reaction valley plot, and is shown in Fig. 1. The minima and deep minima valleys in these plots corresponds to the magicity of projectile/target combinations, and these minima represent the most probable projectile target combinations for fusion of superheavy element.

In the cold reaction valleys of superheavy element $^{302}$120, the probable combinations observed are $^{8}$Be + $^{294}$Lv, $^{10}$Be + $^{292}$Lv, $^{12}$C + $^{290}$Fl, $^{14}$C + $^{288}$Fl, $^{16}$C + $^{286}$Fl, $^{20}$O + $^{282}$Cn, $^{22}$O + $^{280}$Cn, $^{24}$Ne + $^{278}$Ds, $^{26}$Ne + $^{276}$Ds, $^{28}$Mg + $^{274}$Hs, $^{30}$Mg + $^{272}$Hs, $^{32}$Si + $^{270}$Sg, $^{34}$Si + $^{268}$Sg, $^{36}$Si+$^{266}$Sg, $^{38}$S + $^{264}$Rf, $^{40}$S + $^{262}$Rf, $^{42}$S + $^{260}$Rf, etc. The three deep minima are observed in the range 44 < $A_P$ < 60 (region I), 84 < $A_P$ < 100 (region II) and 126 < $A_P$ < 138 (region III) which are due to the magic shell closures of either or both the interacting nuclei. The probable combinations observed in the region I of the cold valley plot are $^{44}$Ar + $^{258}$No, $^{46}$Ar + $^{256}$No, $^{48}$Ca + $^{254}$Fm, $^{50}$Ca + $^{252}$Fm, $^{52}$Ca + $^{250}$Fm, $^{54}$Ti + $^{248}$Cf, $^{56}$Ti + $^{246}$Cf, $^{58}$Cr + $^{244}$Cm, $^{60}$Cr + $^{242}$Cm; in region II the combinations are $^{84}$Se + $^{218}$Rn, $^{86}$Se + $^{216}$Rn, $^{88}$Kr + $^{214}$Po, $^{90}$Kr + $^{212}$Po, $^{92}$Sr + $^{210}$Pb, $^{94}$Sr + $^{208}$Pb, $^{96}$Sr+$^{206}$Pb, $^{98}$Zr+$^{204}$Hg; and in region III the combinations are $^{124}$Sn+$^{178}$Yb, $^{126}$Sn+$^{176}$Yb, $^{128}$Sn+$^{174}$Yb, $^{130}$Te+$^{172}$Er.

In region I, the minima at $^{46}$Ar + $^{256}$No are due to the neutron shell closure N=28 of $^{46}$Ar and the minima at $^{48}$Ca + $^{254}$Fm, $^{50}$Ca + $^{252}$Fm, $^{52}$Ca + $^{250}$Fm are due to the presence of doubly or near doubly magic nuclei Ca. The combinations $^{92}$Sr + $^{210}$Pb, $^{94}$Sr + $^{208}$Pb, $^{96}$Sr +$^{206}$Pb in the region II is because of doubly or near doubly magic nuclei Pb and in region III, the combinations $^{124}$Sn+$^{178}$Yb, $^{126}$Sn + $^{176}$Yb, $^{128}$Sn +$^{174}$Yb, $^{130}$Te+$^{172}$Er, $^{132}$Te + $^{170}$Er were observed due to near doubly magic nuclei Sn and Te and make these systems as suitable projectile-target combinations for the synthesis of super heavy nucleus $^{302}$120. Since the above discussed combinations lie in the cold valleys, they are the optimal cases of binary splitting and hence can be identified as the optimal projectile-target combinations for the synthesis of super heavy element, with considerations to the nature of interaction barrier, potential pocket and the probability of CN formation.

For all the optimal projectile-target combinations identified in the cold valley of super heavy $^{302}$120 nucleus, using Coulomb and proximity potential as the scattering potential we have studied the interaction barriers (scattering potential energy curve) for the fusion of projectile and target against the distance between the centers of the colliding nuclei and the corresponding barrier height $V_B$, the barrier radius $R_B$ and the quasi-fission barrier $B_{qf}$ (depth of the potential well in the nucleus-nucleus interaction) is noted with $\ell = 0$ and the values are given in Table I. While analyzing the interaction barrier, it was found that barrier height $V_B$ is increasing and quasi-fission barrier is decreasing with increasing atomic number of the projectile.

As the part of a systematic study for predicting the most suitable projectile-target combination for heavy ion fusion experiments in the synthesis of $^{302}$120, initially, considered the projectile target combinations $^{126}$Sn + $^{176}$Yb, $^{128}$Sn +$^{174}$Yb, $^{130}$Te + $^{172}$Er, $^{132}$Te + $^{170}$Er found in deep region III of the cold valleys of $^{302}$120. The interaction barrier against the distance between the centers of the projectile and target for the above four combinations are shown in Fig. 2. But, while observing Fig. 2, it is clear that the quasi-fission barrier that are to be appreciable for the fusion to takes place are shallow, in all the four cases and hence cannot be used as a suitable projectile-target combination for heavy ion fusion reactions. Moreover, the projectiles are comparatively heavy, systems are nearly symmetric, and those systems leading to SHE $^{302}$120, the quasi-fission and deep inelastic scattering compete with the fusion process with reduced probability of forming CN. So combinations in region III are not favorable for fusion.

While analyzing the interaction barriers for the rest of the combinations in the cold valley, it is observed that the potential pocket is observed only for the combinations up to $^{102}$Zr+$^{200}$Hg system. The barrier height $V_B$ and barrier radius $R_B$ and the potential pocket for the $^{102}$Zr+$^{200}$Hg system are 318.847MeV, 13.083fm and 478eV respectively. It is noted that, from the interaction barrier, for the combinations $^{84}$Se + $^{218}$Rn, $^{86}$Se + $^{216}$Rn, $^{88}$Kr + $^{214}$Po, $^{90}$Kr + $^{212}$Po (shown in Fig. 3), $^{92}$Sr + $^{210}$Pb, $^{94}$Sr + $^{208}$Pb, $^{96}$Sr + $^{206}$Pb, $^{98}$Zr+$^{204}$Hg (shown in Fig. 4) found in region II, the potential pocket is small compared with that for the combinations $^{44}$Ar + $^{258}$No, $^{46}$Ar + $^{256}$No, $^{48}$Ca + $^{254}$Fm, $^{50}$Ca + $^{252}$Fm, (shown in Fig. 5); $^{52}$Ca + $^{250}$Fm, $^{54}$Ti + $^{248}$Cf, $^{56}$Ti + $^{246}$Cf, $^{58}$Cr + $^{244}$Cm (shown in Fig. 6) found in region I. The potential pocket is appreciable in the cases of combinations found in the cold valley region I and that of combinations in region II . So, combinations in first deep region and second deep region can be identified as the most probable projectile-target combinations for the fusion. Excitation energy of the combinations in region I near and above barrier is comparatively higher than that in region II because combinations in region I are more asymmetric. So we can take combinations in region I as favorable for hot fusion reaction and the combinations in region II is favorable for cold fusion reactions.

Further, in an attempt to predict more suitable projectile-target, which are having good potential pockets, we have considered the projectiles and targets having comparatively large half-lives and the systems $^{20}$O + $^{282}$Cn, $^{36}$Si + $^{266}$Sg, $^{40}$S + $^{262}$Rf, $^{62}$Fe + $^{240}$Pu, $^{64}$Fe + $^{238}$Pu, $^{66}$Fe + $^{236}$Pu, $^{68}$Ni + $^{234}$U, $^{70}$Ni + $^{232}$U, $^{72}$Ni + $^{230}$U, $^{74}$Zn + $^{228}$Th are found in the other cold valley are also feasible for fusion experiments. It is noted that as the reaction asymmetry increases, excitation energy also increases. The excitation energy near and above the barrier for the combinations $^{20}$O + $^{282}$Cn, $^{36}$Si + $^{266}$Sg, $^{40}$S + $^{262}$Rf, which are more mass asymmetric than the combinations in region I, is comparatively higher and thus have less survival probability to form a evaporation residue and these combinations are not at all favorable for fusion. Based on the two simple arguments of reaction asymmetry and excitation energy, the combinations $^{20}$O + $^{282}$Cn, $^{36}$Si + $^{266}$Sg, $^{40}$S + $^{262}$Rf are not a promising choice for an attempt to synthesize the element $^{302}$120.

We have evaluated the capture cross sections as a function of center of mass energy (excitation function) for the predicted combinations using Wong formula, approximated Wong formula, and Glas and Mosel formula. The corresponding excitation function $\sigma_{capture}$ versus $E_{CM}$ are given in the upper panel of Figs. 14-25. It is found that the Wong formula is in good agreement with the Glas and Mosel formula.

In order to find the fusion cross section for these combinations of superheavy $^{302}$120, the knowledge of probability of compound nucleus formation $P_{CN}$ is a must and is determined using the equation (21) and are plotted against excitation energy of compound nucleus which are shown in Figs. 7-12. It is found that, $P_{CN}$ for the combinations in region II (less asymmetric cold combinations) is very small as compared to combinations in region I (more asymmetric hot combination). It is found that $P_{CN}$ is larger for more asymmetric hot combination. For a given excitation energy (E*=35 MeV for hot fusion reaction, E* = 5 MeV for cold fusion reaction), it can also be seen from the Fig. 13 that $P_{CN}$ is larger for the more asymmetric hot fusion combinations. Fusion probability increases with increasing quasi-fission barrier height (potential pocket) and the incident energies in the reactions leading to the SHE.

Near and above the barrier, using the values of $P_{CN}$ for each center of mass energy, fusion cross section is computed by using the equation $\sigma_{fusion} = \sigma_{capture} \times P_{CN}$ for the above systems and the corresponding fusion excitation functions ($\sigma_{fusion}$ versus $E_{CM}$ plots) are shown in upper panels of Figs.14-25. From the plots it can be seen that computed fusion cross section for combinations in the first deep region $44 < A_P < 58$ is in the order of micro barn, which is higher than in the second region $84 < A_P < 100$, where the fusion cross section is in the order of pico barn and for the combinations in region $60 < A_P < 82$, fusion cross section is in between that of region I and II. The fusion cross sections for more asymmetric (and "hotter") fusion reactions found to be higher than symmetric "colder" combinations. The combinations with large fusion cross sections are the ones which are more asymmetric and having more potential pockets (quasi-fission barrier).

For the de-excitation stage, the survival probability $W_{sur}$ is calculated by using the_formalism discussed in sub section 3.1. We calculated the evaporation residue cross sections for the *2n, 3n, 4n* and *5n* evaporation channels for the hot combinations, and *1n* evaporation channel for the cold combinations and plotted in the lower panels of Figs. 14-25. The predictions of the maximum value of the ER cross sections of all the probable combinations are presented in Table 2. The calculated cross section for *3n* channel (84.21fb) and *4n* channel (15.67fb) is larger for the reaction $^{48}$Ca+$^{254}$Fm and *2n* cross section (325.21fb) is larger for $^{68}$Ni+$^{234}$U. Predicted *1n* cross section (shown in lower panel of figures 22-25) for cold fusion combinations is too small as compared to hot fusion combinations. So to synthesize element Z=120, our study reveals that, hot fusion reaction is preferable.

As shown in Fig. 14(c) and 14(d), maximum ER cross section (*2n* and *3n*) for $^{46}$Ar+$^{256}$No is found to be higher than $^{44}$Ar+$^{258}$No. Also among the combinations $^{48}$Ca+$^{254}$Fm, $^{50}$Ca+$^{252}$Fm and $^{52}$Ca+$^{250}$Fm (Fig. 15(c), 15(d) and 16(c)), our result shows that $\sigma_{ER}$ is larger for $^{48}$Ca +$^{254}$Fm and is the most favorable projectile target combination. The ER cross sections for $^{54}$Ti+$^{248}$Cf (Fig.16 (d)) are higher than $^{56}$Ti+$^{246}$Cf (Fig 17(c)). Among the combinations $^{58}$Cr+$^{244}$Cm (Fig. 17(d)), $^{60}$Cr+$^{242}$Cm (Fig18(c)), latter reaction is more favorable to *2n* channel and former reaction is favorable to *3n* and *4n* channel. Similarly for the reactions $^{62}$Fe+$^{240}$Pu (Fig.18 (d)), $^{64}$Fe+$^{238}$Pu (Fig19(c)) and $^{66}$Fe+$^{236}$Pu (Fig.19 (d)), $\sigma_{ER}$ (*2n*) is more for $^{64}$Fe+$^{238}$Pu whereas $\sigma_{ER}$ (*3n* and *4n*) is more for $^{62}$Fe+$^{240}$Pu. The ER cross sections of $^{68}$Ni+$^{234}$U (Fig. 20(c)) are more than $^{70}$Ni+$^{232}$U (Fig.20 (d)) and $^{72}$Ni+$^{230}$U (Fig. 21(c)) systems. Based on the arguments of quasi-fission barrier, reaction asymmetry, excitation energy, probability of compound nucleus formation and survival probability, the most promising choice for an attempt to synthesize element $^{302}$120 are $^{44}$Ar + $^{258}$No, $^{46}$Ar + $^{256}$No, $^{48}$Ca + $^{254}$Fm, $^{50}$Ca + $^{252}$Fm, $^{54}$Ti + $^{248}$Cf, $^{58}$Cr + $^{244}$Cm, $^{62}$Fe + $^{240}$Pu, $^{64}$Fe + $^{238}$Pu, $^{68}$Ni + $^{234}$U, $^{70}$Ni + $^{232}$U, $^{72}$Ni + $^{230}$U, $^{74}$Zn + $^{228}$Th.

In order to check the consistency of our calculations, we have evaluated the capture, fusion, and evaporation residue cross sections of four reactions, $^{54}$Cr+$^{248}$Cm, $^{58}$Fe+$^{244}$Pu, $^{64}$Ni+$^{238}$U and $^{50}$Ti+$^{249}$Cf, for which several theoretical studies and attempt to produce the element Z=120 were done. Fig. 26 shows the predicted capture, fusion and ER cross section for $^{54}$Cr+$^{248}$Cm and $^{58}$Fe+$^{244}$Pu systems. The calculated

*3n* and *4n* channel ER cross sections for the reaction $^{54}Cr+^{248}Cm$ (Fig. 26(c)) is 15.4fb and 5.04fb respectively and for $^{58}Fe+^{244}Pu$ reaction (Fig. 26(d)), it is found to be 6.12fb and 2.78fb respectively. The excitation functions for $^{64}Ni+^{238}U$ and $^{50}Ti+^{249}Cf$ shown in Fig. 27. The *3n* and *4n* channel maximum ER cross section for $^{64}Ni+^{238}U$ (Fig. 27(c)) are 7.81fb and 1.49fb respectively, and that of $^{50}Ti+^{249}Cf$ (Fig. 27(d)) are 102.92fb and 2.264fb respectively. We have observed that the more mass asymmetric $^{50}Ti+^{249}Cf$ is more favorable to synthesize element Z=120 because *3n* and *4n* evaporation residue cross section are much larger than the maximal values for the other reactions. To compare the results with other models, we listed the maximum ER cross sections (*3n* and *4n*) for the combinations $^{54}Cr+^{248}Cm$, $^{58}Fe+^{244}Pu$, $^{64}Ni+^{238}U$ and $^{50}Ti+^{249}Cf$ in Tables 3 and 4. The calculated maximum *3n* and *4n* evaporation residue cross sections for the above mentioned combinations are of the same order with other theoretical models and $^{50}Ti+^{249}Cf$ is found to the more feasible combination in all the theoretical studies, which very well establishes the reliability of our work. Note that experimental upper limit for above reactions $^{58}Fe+^{244}Pu$, $^{64}Ni+^{238}U$ and $^{54}Cr+^{248}Cm$ had been established at 400fb[2], 90fb[39], 560fb[40,41] respectively and an attempt was done using the reactions $^{50}Ti+^{249}Cf$ [85]. Recently Hofmann et al., [86] measured the cross section for the three event chain observed in the reaction $^{54}Cr+^{248}Cm$ tentatively assigned to $^{299}120$ is $0.58^{+1.34}_{-0.48}$ pb.

## IV. CONCLUSIONS

Probable target- projectile combinations for the super heavy element $^{302}120$ have been identified from the cold reaction valleys. We have calculated the interaction barriers for the fusion of all the projectile-target combinations identified in the cold valleys of super heavy $^{302}120$ nucleus, against the distance between the centers of the projectile and target by taking Coulomb and proximity potential as the scattering potential. Near and above the barrier, the total capture, fusion and ER cross sections for all the systems also have been calculated. The systems $^{44}Ar + ^{258}No$, $^{46}Ar + ^{256}No$, $^{48}Ca +^{254}Fm$, $^{50}Ca + ^{252}Fm$, $^{52}Ca + ^{250}Fm$, $^{54}Ti + ^{248}Cf$, $^{56}Ti + ^{246}Cf$, $^{58}Cr + ^{244}Cm$ in the deep region I of cold valley, and the systems $^{60}Cr + ^{242}Cm$, $^{62}Fe + ^{240}Pu$, $^{64}Fe + ^{238}Pu$, $^{66}Fe + ^{236}Pu$, $^{68}Ni + ^{234}U$, $^{70}Ni + ^{232}U$, $^{72}Ni + ^{230}U$, $^{74}Zn + ^{228}Th$ in the cold valleys are identified as the better projectile target combinations for the synthesis of $^{302}120$. While considering the nature of quasi-fission barrier height and half-lives of colliding nuclei, mass asymmetry, probability of compound nucleus formation, survival probability and excitation energy, the systems $^{44}Ar + ^{258}No$, $^{46}Ar + ^{256}No$, $^{48}Ca + ^{254}Fm$, $^{50}Ca + ^{252}Fm$, $^{54}Ti + ^{248}Cf$, $^{58}Cr + ^{244}Cm$, $^{62}Fe + ^{240}Pu$, $^{64}Fe + ^{238}Pu$, $^{68}Ni + ^{234}U$, $^{70}Ni + ^{232}U$, $^{72}Ni + ^{230}U$, $^{74}Zn + ^{228}Th$ give maximum probability for the synthesis of super heavy nucleus $^{302}120$. The computed ER cross section for $^{54}Cr+^{248}Cm$, $^{58}Fe+^{244}Pu$, $^{64}Ni+^{238}U$ and $^{50}Ti+^{249}Cf$ combinations are compared with experimental data and other theoretical models, and all models predicted the maximum cross section for the combination $^{50}Ti+^{249}Cf$, which proves reliability of our work. Through our extensive study, we predict several promising possibilities for the synthesis of SHE $^{302}120$.

-------------------------------------------------

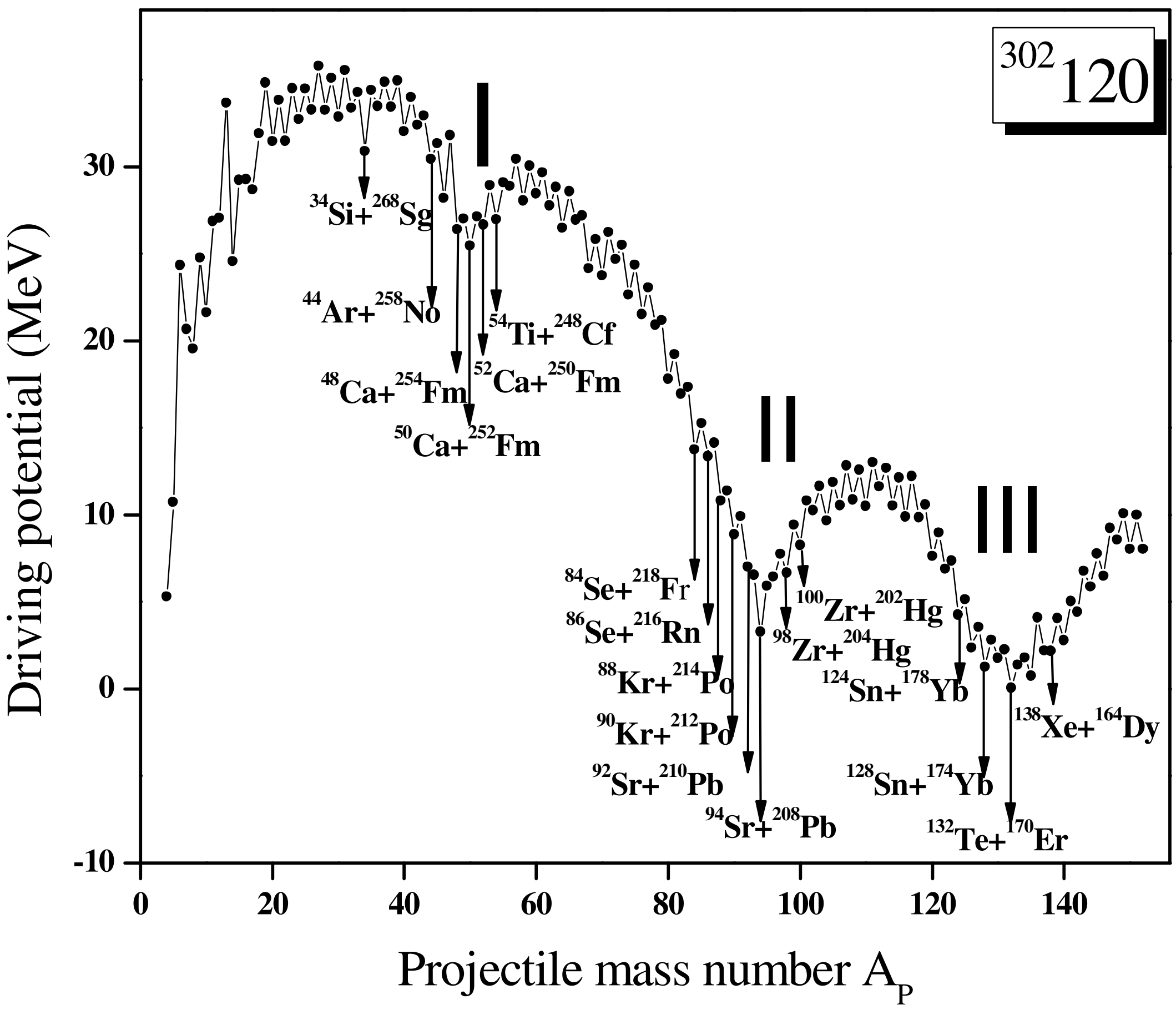

FIG 1. Cold reaction valley plot of superheavy nuclei $^{302}120$

TABLE 1. Barrier height $V_B$, barrier radius $R_B$ and quasi-fission barrier height for the systems in the cold reaction valleys of $^{302}120$ nuclei taking Coulomb and proximity potential as interacting barrier.

| Reaction | Barrier height $V_B$ (MeV) | Barrier radius $R_B$ (fm) | Quasi-fission barrier $B_{qf}$ (MeV) | Reaction | Barrier height $V_B$ (MeV) | Barrier radius $R_B$ (fm) | Quasi-fission barrier $B_{qf}$ (MeV) |
|---|---|---|---|---|---|---|---|
| $^{8}Be+^{294}Lv$ | 51.214 | 12.300 | 19.786 | $^{56}Ti+^{246}Cf$ | 219.328 | 13.343 | 7.688 |
| $^{10}Be+^{292}Lv$ | 50.279 | 12.430 | 24.283 | $^{58}Cr+^{244}Cm$ | 234.671 | 13.206 | 5.652 |
| $^{12}C+^{290}Fl$ | 74.756 | 12.375 | 17.143 | $^{60}Cr+^{242}Cm$ | 233.940 | 13.269 | 6.210 |
| $^{14}C+^{288}Fl$ | 73.764 | 12.585 | 21.143 | $^{62}Fe+^{240}Pu$ | 248.440 | 13.146 | 4.188 |
| $^{16}C+^{286}Fl$ | 72.897 | 12.708 | 24.648 | $^{64}Fe+^{238}Pu$ | 247.714 | 13.190 | 4.507 |
| $^{20}O+^{282}Cn$ | 95.119 | 12.749 | 21.076 | $^{66}Fe+^{236}Pu$ | 247.050 | 13.300 | 5.341 |
| $^{22}O+^{280}Cn$ | 94.316 | 12.921 | 23.355 | $^{68}Ni+^{234}U$ | 260.650 | 13.163 | 3.513 |
| $^{24}Ne+^{278}Ds$ | 116.278 | 12.837 | 18.055 | $^{70}Ni+^{232}U$ | 259.960 | 13.217 | 4.188 |
| $^{26}Ne+^{276}Ds$ | 115.418 | 12.922 | 19.850 | $^{72}Ni+^{230}U$ | 259.367 | 13.320 | 4.507 |
| $^{28}Mg+^{274}Hs$ | 136.364 | 12.889 | 15.037 | $^{74}Zn+^{228}Th$ | 272.110 | 13.210 | 2.839 |
| $^{30}Mg+^{272}Hs$ | 135.497 | 12.983 | 16.455 | $^{76}Zn+^{226}Th$ | 271.470 | 13.246 | 3.407 |
| $^{32}Si+^{270}Sg$ | 155.440 | 12.943 | 12.533 | $^{78}Ge+^{224}Ra$ | 283.464 | 13.149 | 2.061 |
| $^{34}Si+^{268}Sg$ | 154.551 | 13.012 | 13.940 | $^{80}Ge+^{222}Ra$ | 282.813 | 13.146 | 2.444 |
| $^{36}Si+^{266}Sg$ | 153.748 | 13.103 | 15.150 | $^{82}Ge+^{220}Ra$ | 282.223 | 13.244 | 2.825 |
| $^{38}S+^{264}Rf$ | 172.685 | 13.042 | 11.409 | $^{84}Se+^{218}Rn$ | 293.808 | 13.198 | 1.867 |
| $^{40}S+^{262}Rf$ | 171.860 | 13.128 | 12.530 | $^{86}Se+^{216}Rn$ | 292.808 | 13.216 | 2.163 |
| $^{42}S+^{260}Rf$ | 171.075 | 13.180 | 13.490 | $^{88}Kr+^{214}Po$ | 303.174 | 13.118 | 0.721 |
| $^{44}Ar+^{258}No$ | 189.080 | 13.127 | 11.178 | $^{90}Kr+^{212}Po$ | 302.644 | 13.192 | 0.874 |
| $^{46}Ar+^{256}No$ | 188.282 | 13.176 | 10.216 | $^{92}Sr+^{210}Pb$ | 312.046 | 13.028 | 0.493 |
| $^{48}Ca+^{254}Fm$ | 205.389 | 13.120 | 8.174 | $^{94}Sr+^{208}Pb$ | 311.499 | 13.051 | 0.577 |
| $^{50}Ca+^{252}Fm$ | 204.603 | 13.175 | 9.105 | $^{98}Zr+^{204}Hg$ | 319.803 | 12.992 | 0.294 |
| $^{52}Ca+^{250}Fm$ | 203.873 | 13.259 | 9.867 | $^{100}Zr+^{202}Hg$ | 319.312 | 13.060 | 0.458 |
| $^{54}Ti+^{248}Cf$ | 220.070 | 13.200 | 7.223 | $^{102}Zr+^{200}Hg$ | 318.847 | 13.083 | 0.478 |

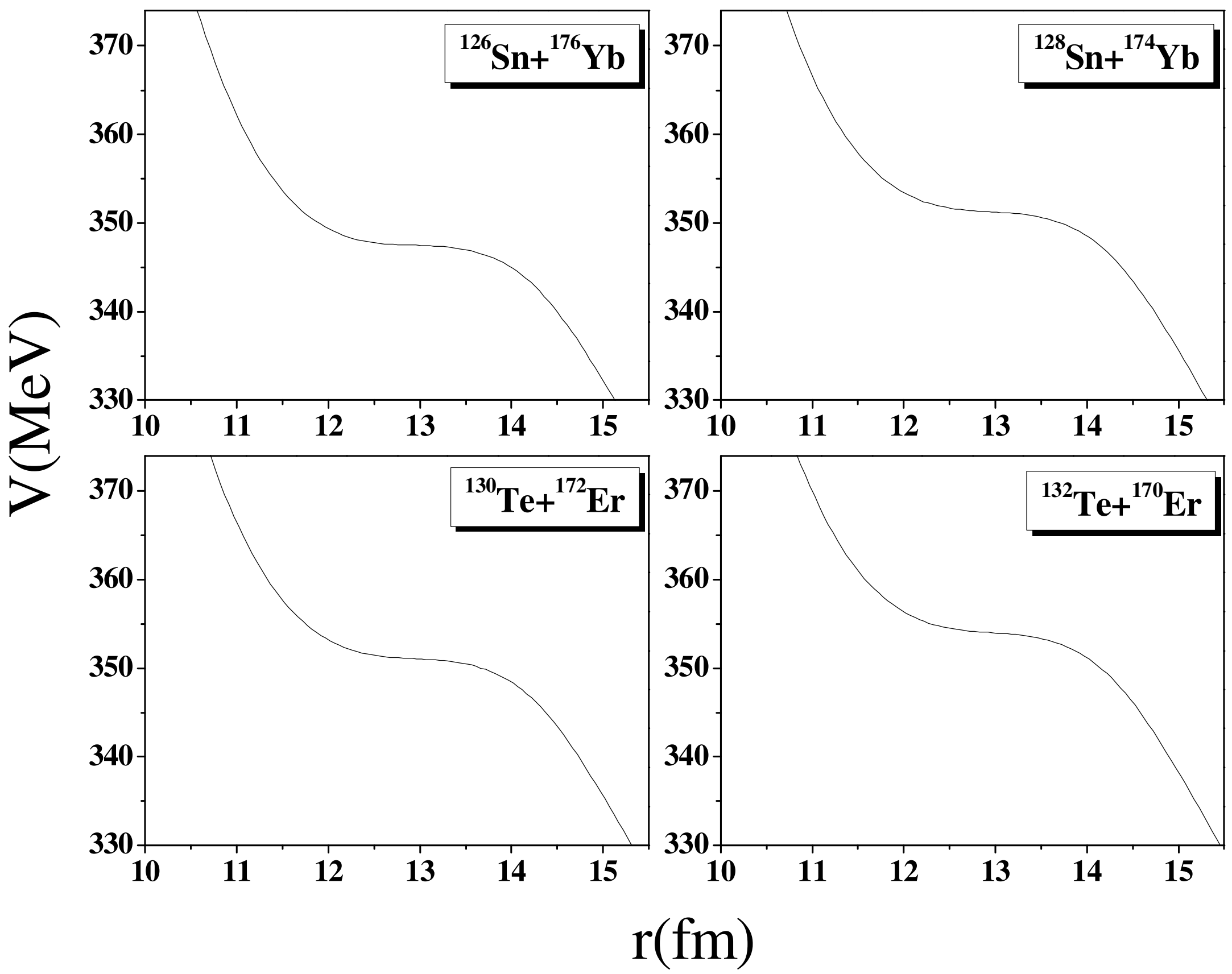

FIG 2. Scattering potential for the reactions of $^{126}Sn$ + $^{176}Yb$, $^{128}Sn$ + $^{174}Yb$, $^{130}Te$ + $^{172}Er$ and $^{132}Te$ + $^{170}Er$ systems.

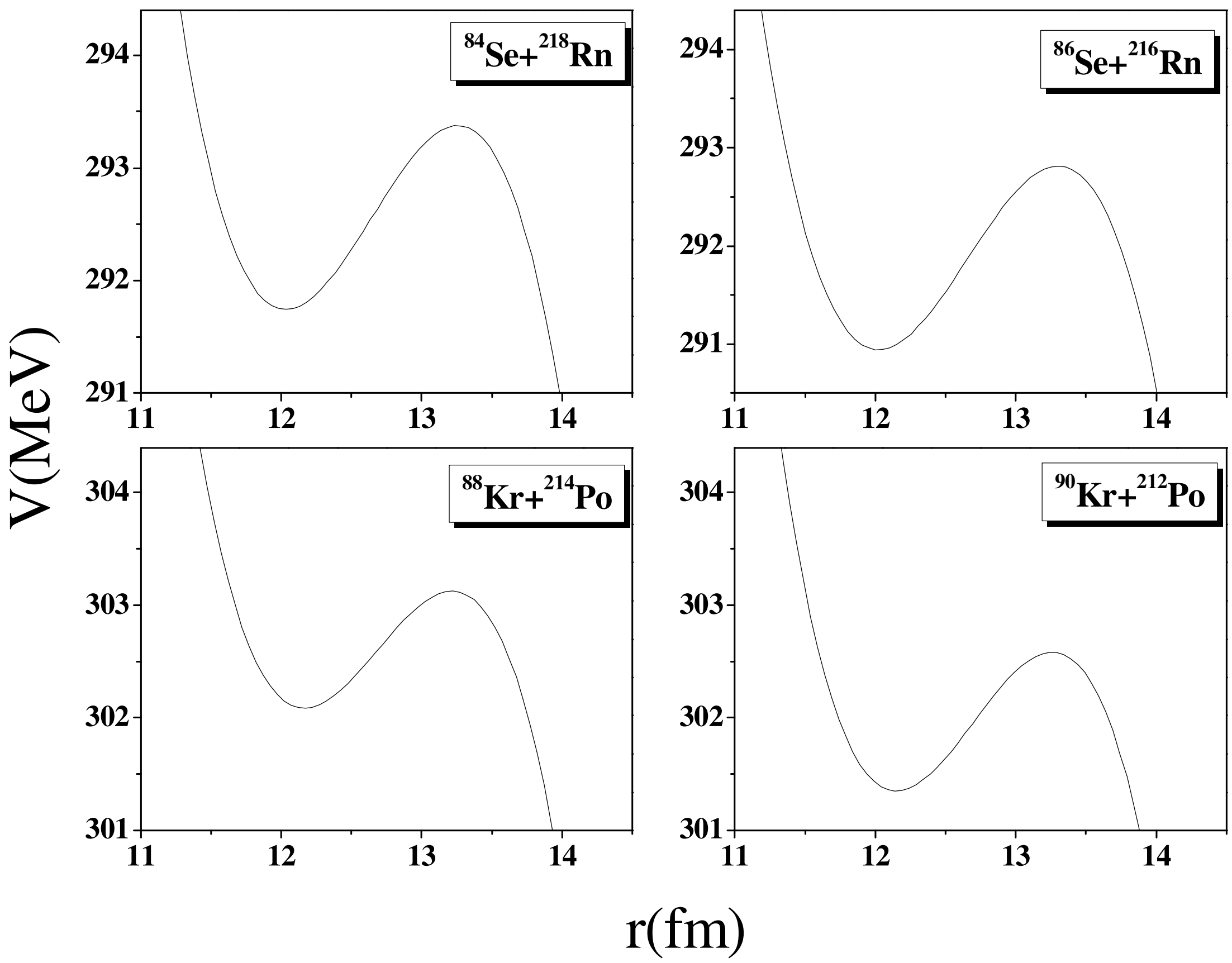

FIG 3. Scattering potential for the reactions of $^{84}Se+^{218}Rn$, $^{86}Se+^{216}Rn$, $^{88}Kr+^{214}Po$ and $^{90}Kr+^{212}Po$ systems.

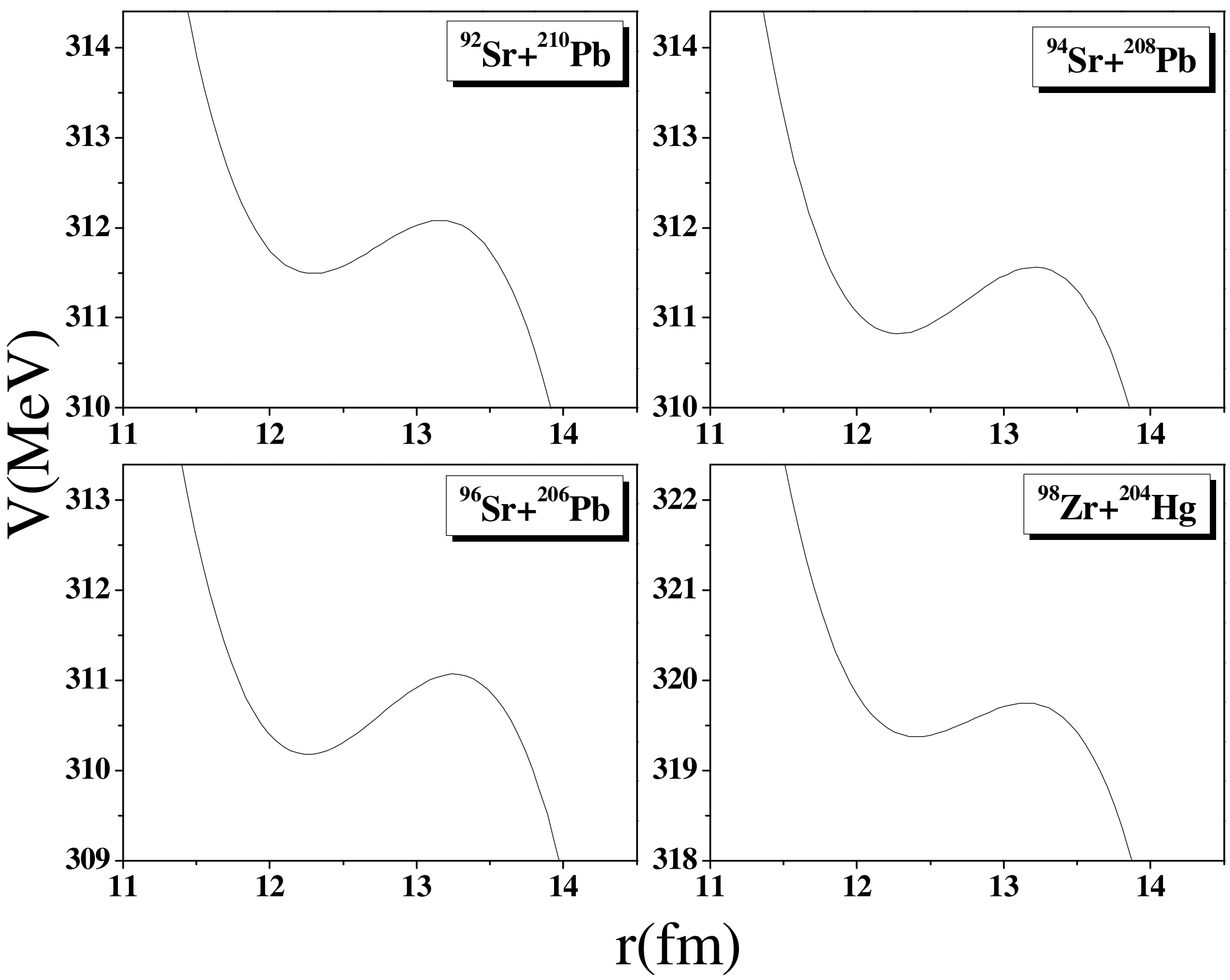

FIG 4. Scattering potential for the reactions of $^{92}$Sr+$^{210}$Pb, $^{94}$Sr+$^{208}$Pb, $^{96}$Sr+$^{206}$Pb and $^{98}$Zr+$^{204}$Hg systems.

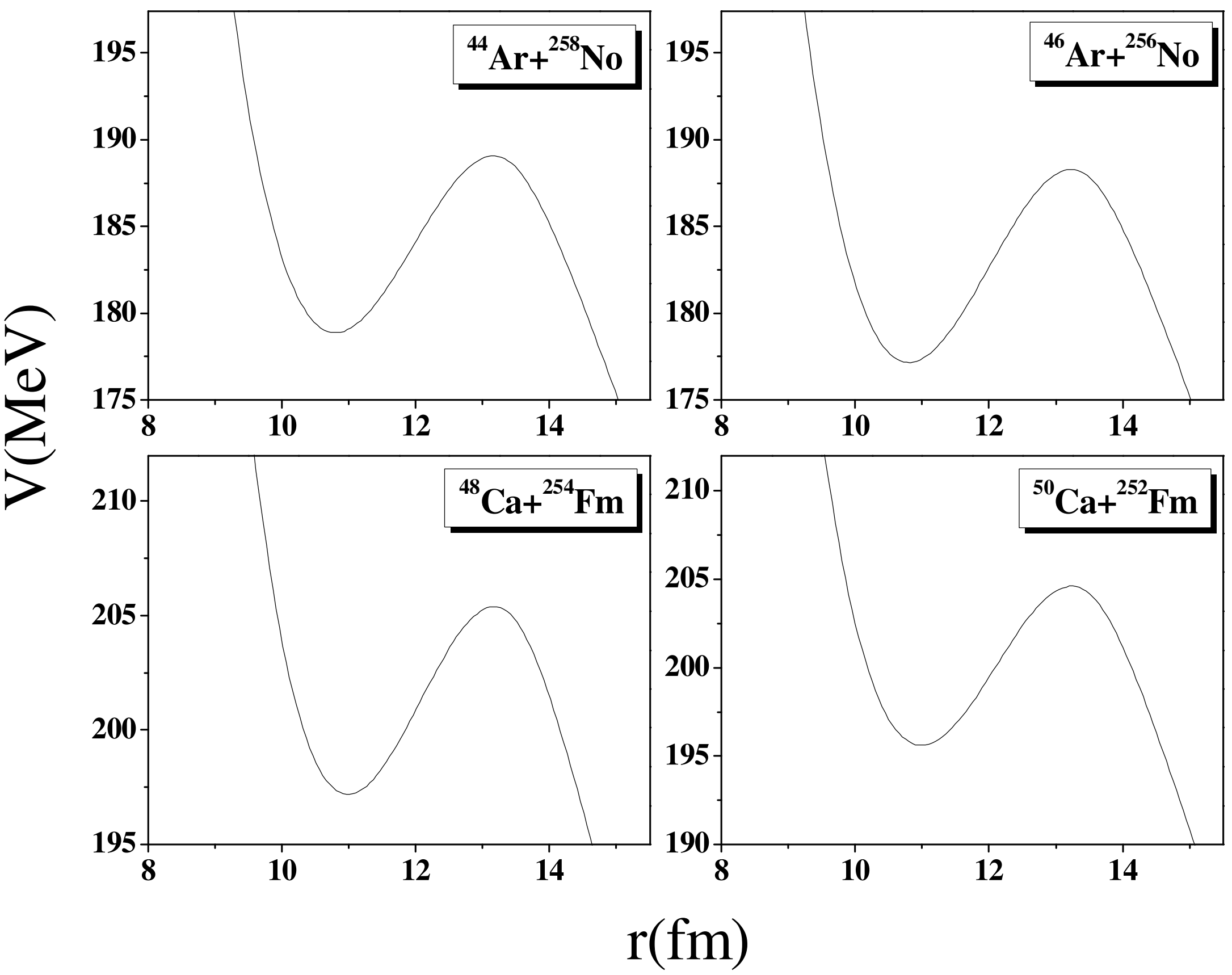

FIG 5. Scattering potential for the reactions of $^{44}$Ar+$^{258}$No, $^{46}$Ar+$^{256}$No, $^{48}$Ca+$^{254}$Fm and $^{50}$Ca+$^{252}$Fm systems.

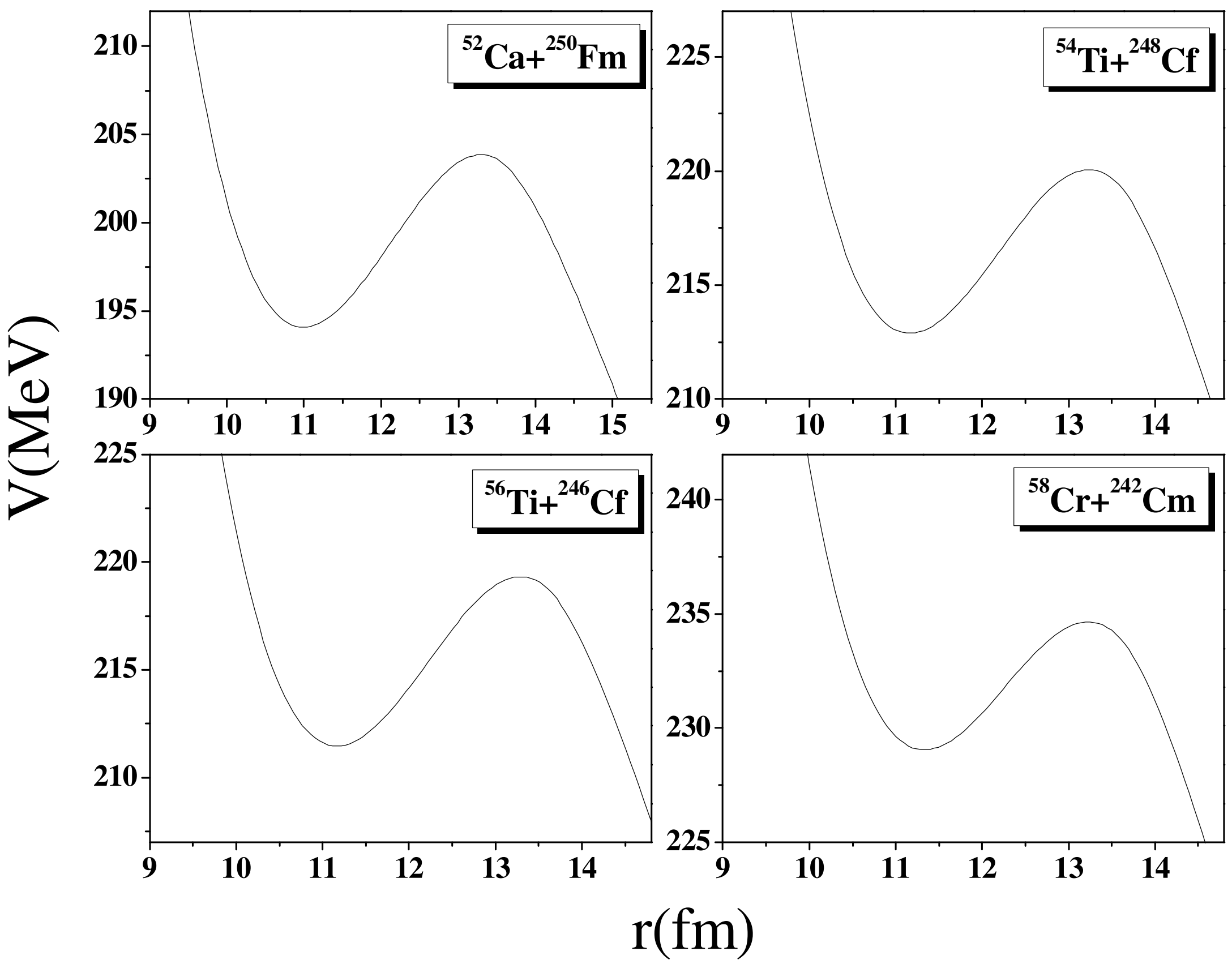

FIG 6. Scattering potential for the reactions of $^{52}$Ca+$^{250}$Fm, $^{54}$Ti+$^{248}$Cf, $^{56}$Ti+$^{246}$Cf and $^{58}$Cr+$^{244}$Cm systems

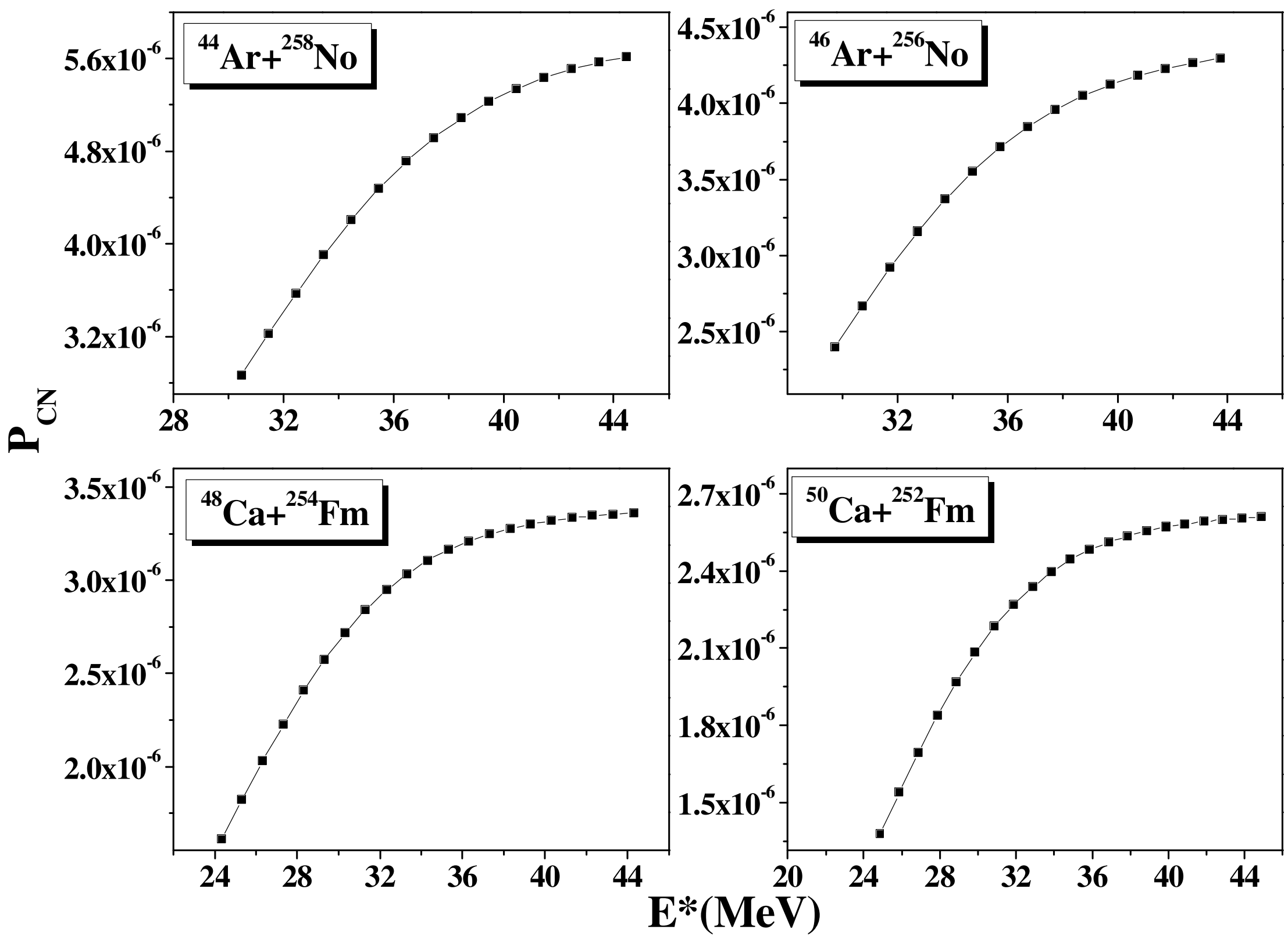

FIG 7. The plot of $P_{CN}$ vs. $E^*$ in MeV for the reactions of $^{44}$Ar+$^{258}$No, $^{46}$Ar+$^{256}$No, $^{48}$Ca+$^{254}$Fm and $^{50}$Ca+$^{252}$Fm systems.

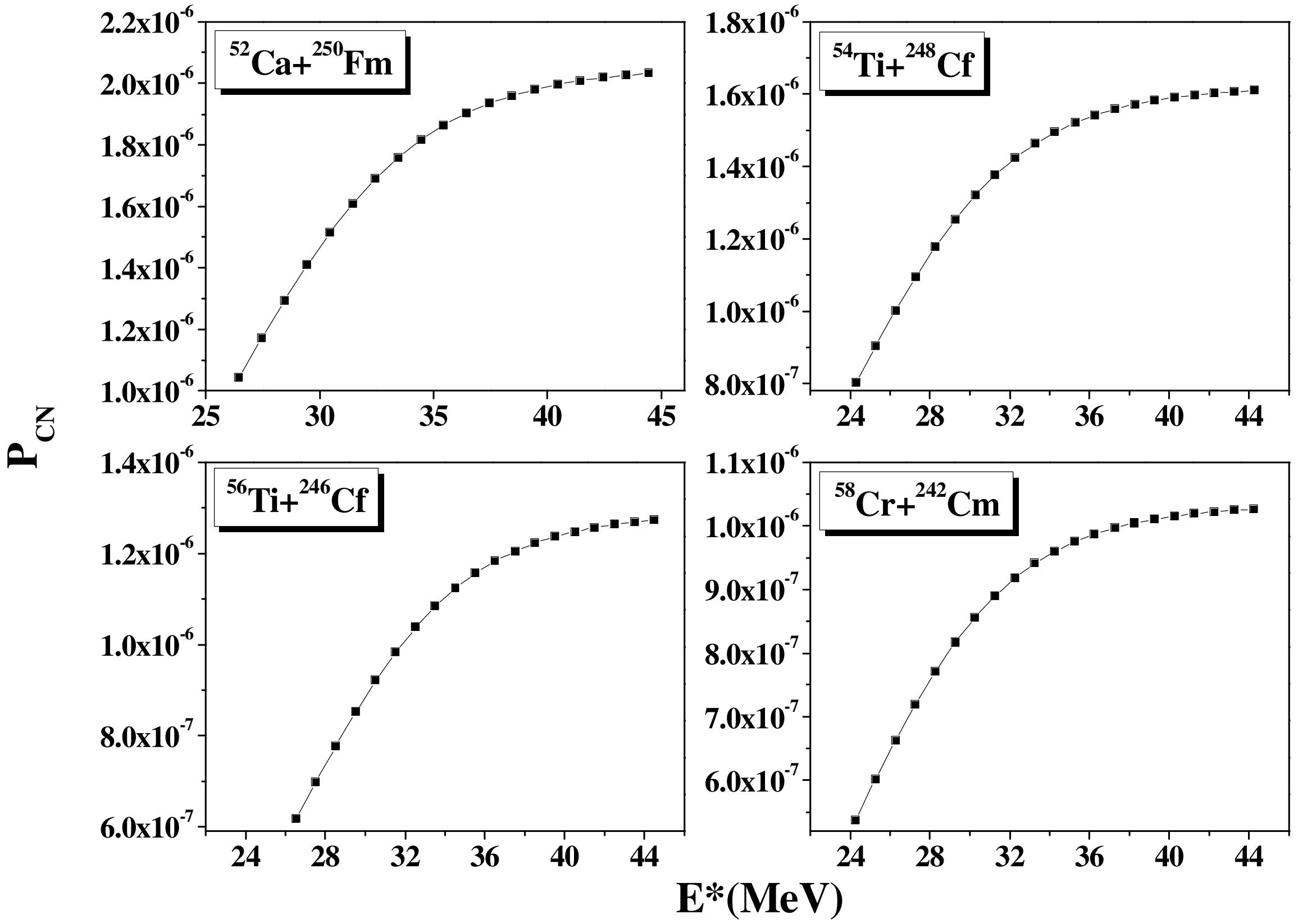

FIG 8. The plot of $P_{CN}$ vs. $E^*$ in MeV for the reactions of $^{52}$Ca+$^{250}$Fm, $^{54}$Ti+$^{248}$Cf, $^{56}$Ti+$^{246}$Cf, $^{58}$Cr+$^{244}$Cm systems

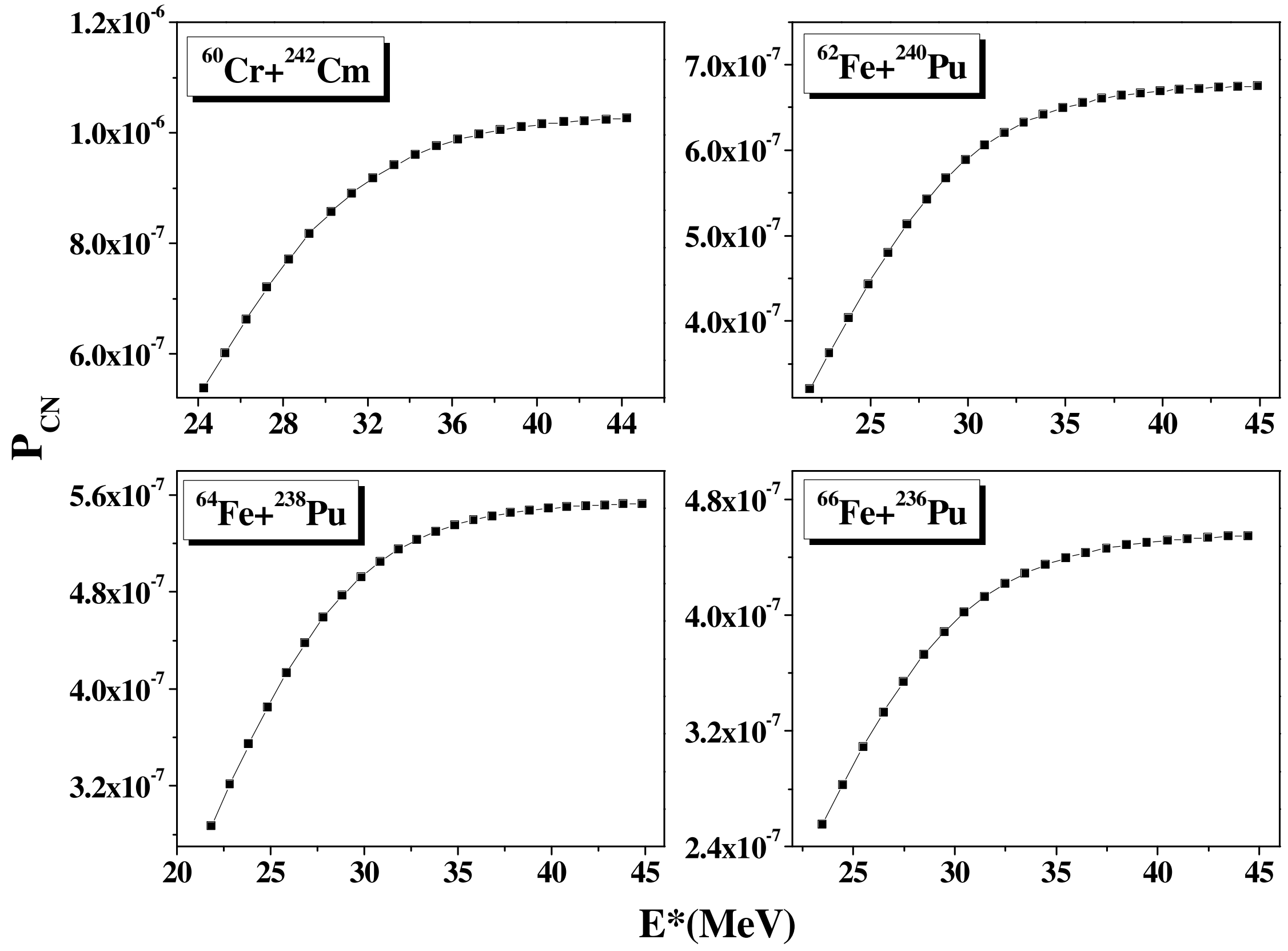

FIG 9.The plot of $P_{CN}$ vs. $E^*$ in MeV for the reactions of $^{60}$Cr + $^{242}$Cm, $^{62}$Fe + $^{240}$Pu, $^{64}$Fe + $^{238}$Pu and $^{66}$Fe + $^{236}$Pu systems.

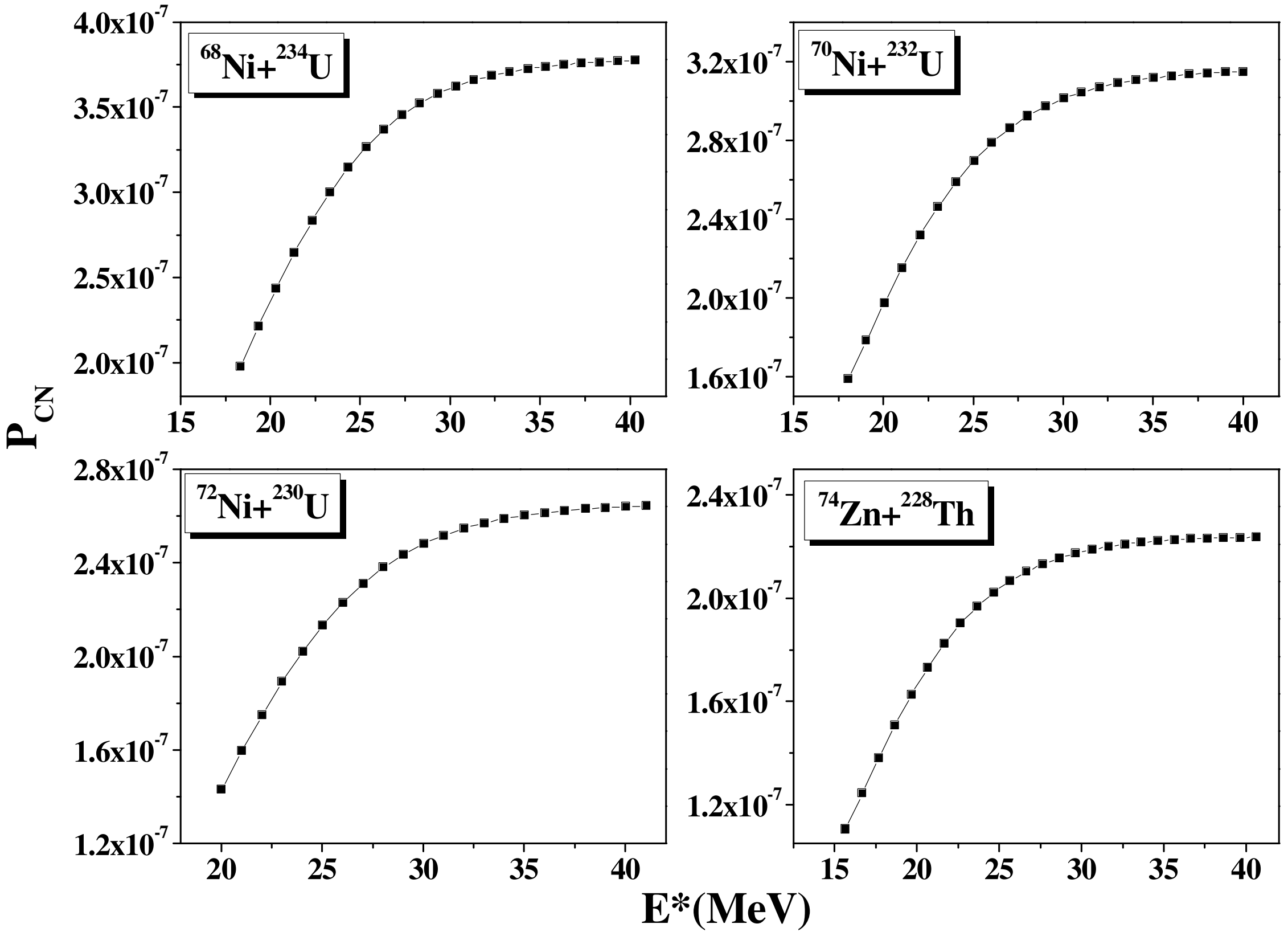

FIG 10.The plot of $P_{CN}$ vs. $E^*$ in MeV for the reactions of $^{68}$Ni + $^{234}$U, $^{70}$Ni + $^{232}$U, $^{72}$Ni + $^{230}$U and $^{74}$Zn + $^{228}$Th systems.

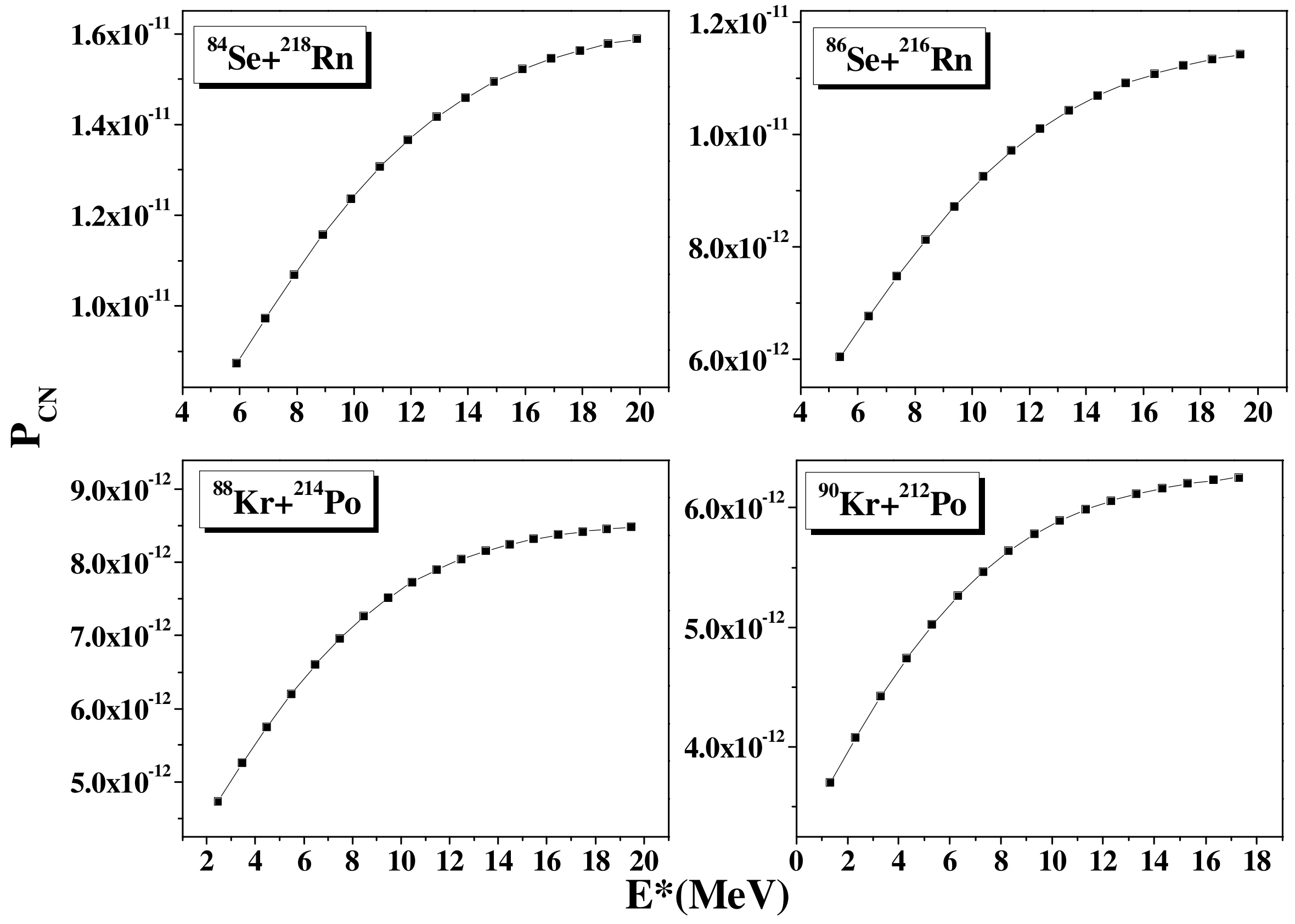

FIG 11.The plot of $P_{CN}$ vs. $E^*$ in MeV for the reactions of $^{84}$Se+$^{218}$Rn, $^{86}$Se+$^{216}$Rn, $^{88}$Kr+$^{214}$Po and $^{90}$Kr+$^{212}$Po systems

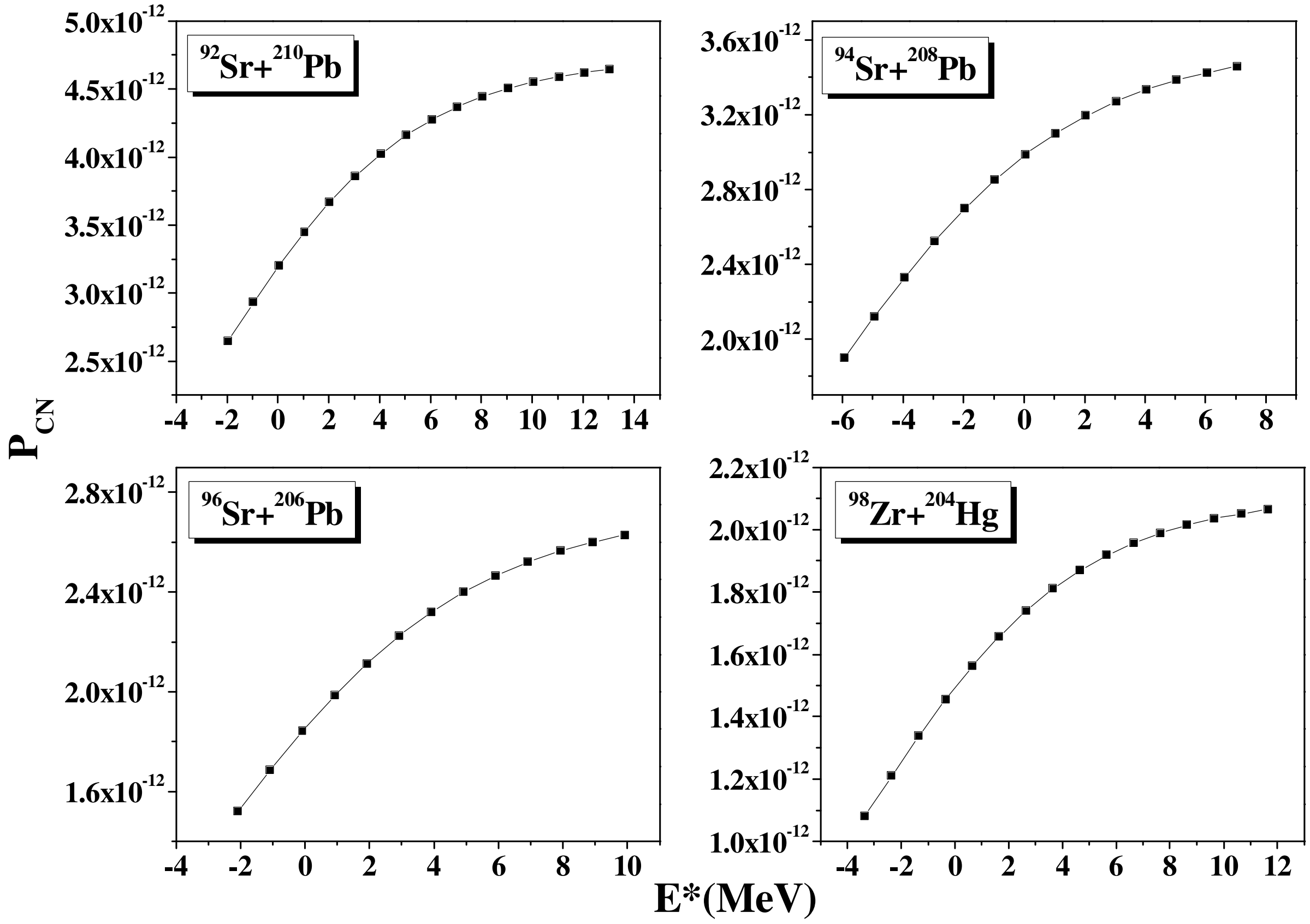

FIG 12.The plot of $P_{CN}$ vs. $E^*$ in MeV for the reactions of $^{92}$Sr+$^{210}$Pb, $^{94}$Sr+$^{208}$Pb, $^{96}$Sr+$^{206}$Pb and $^{98}$Zr+$^{204}$Hg systems.

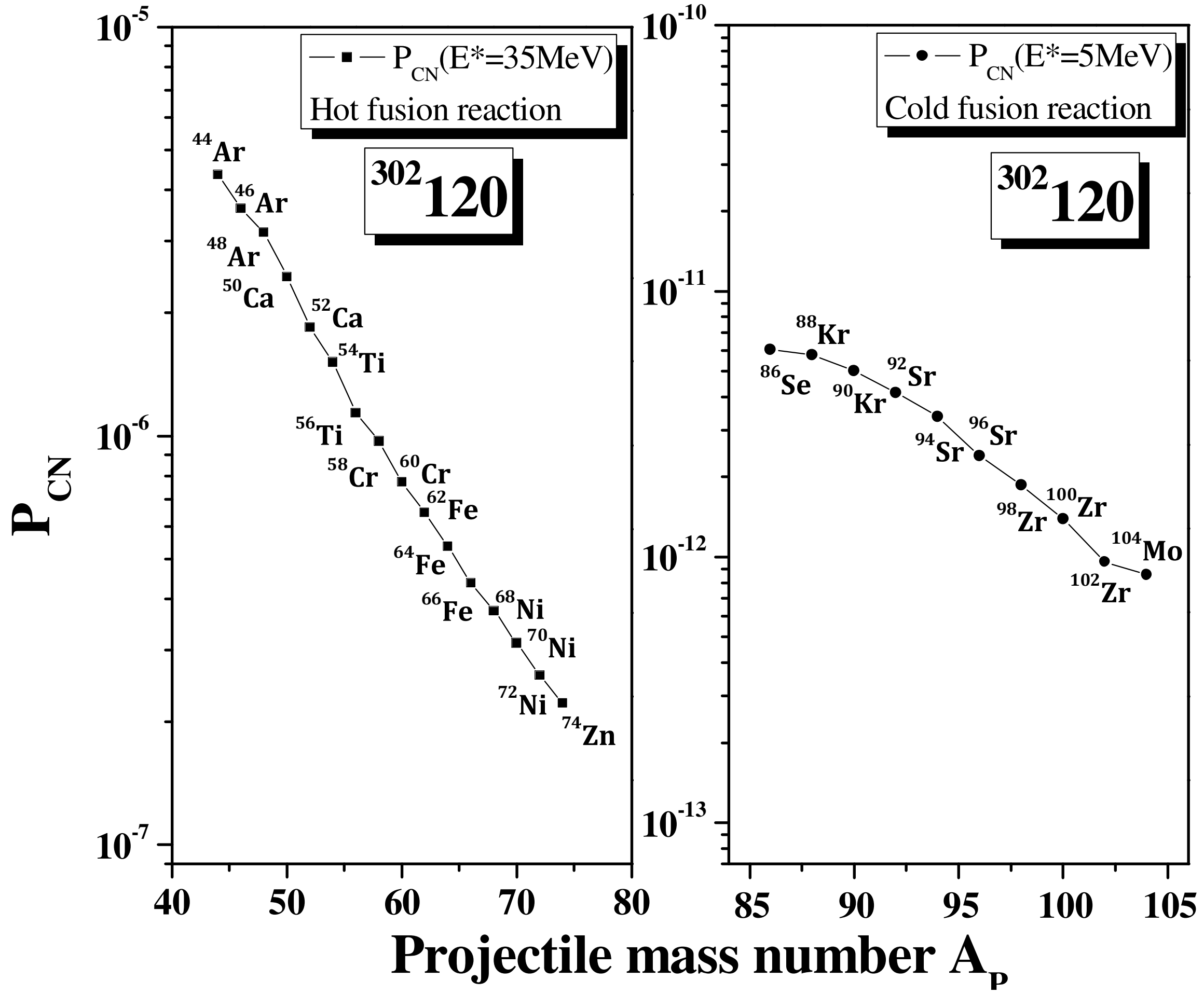

FIG 13. Plot of $P_{CN}$ for hot fusion (E*=35MeV) and cold fusion (E*=5MeV) reactions against mass number of the projectile $A_P$ for SHE $^{302}$120.

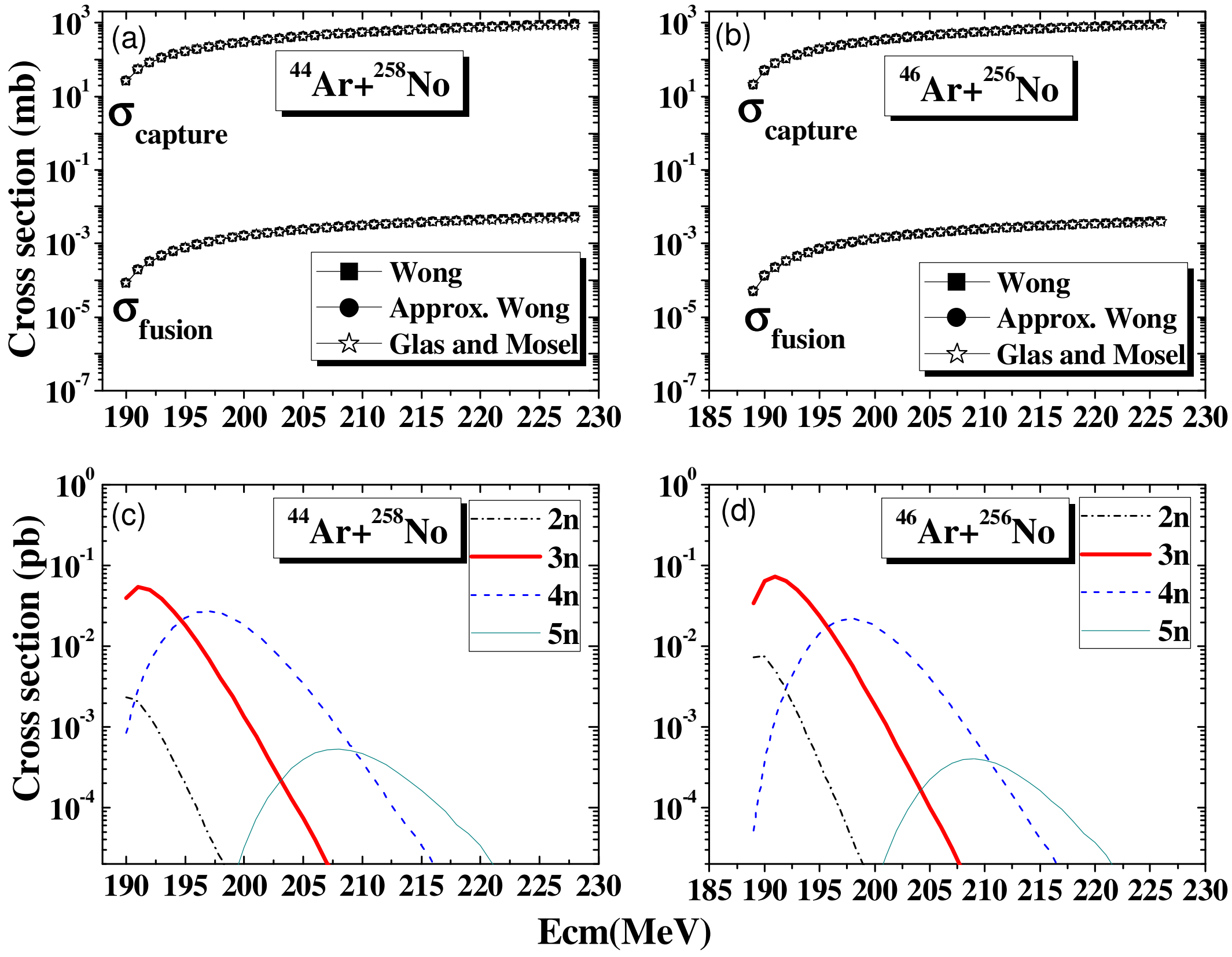

FIG 14. Plots of capture ($\sigma_{capture}$) and fusion ($\sigma_{fusion}$) cross sections in *mb* (upper panel) and evaporation residue cross section in *pb* (lower panel) vs. center of mass energy *Ecm* in MeV for the reactions of $^{44}$Ar + $^{258}$No and $^{46}$Ar + $^{256}$No systems

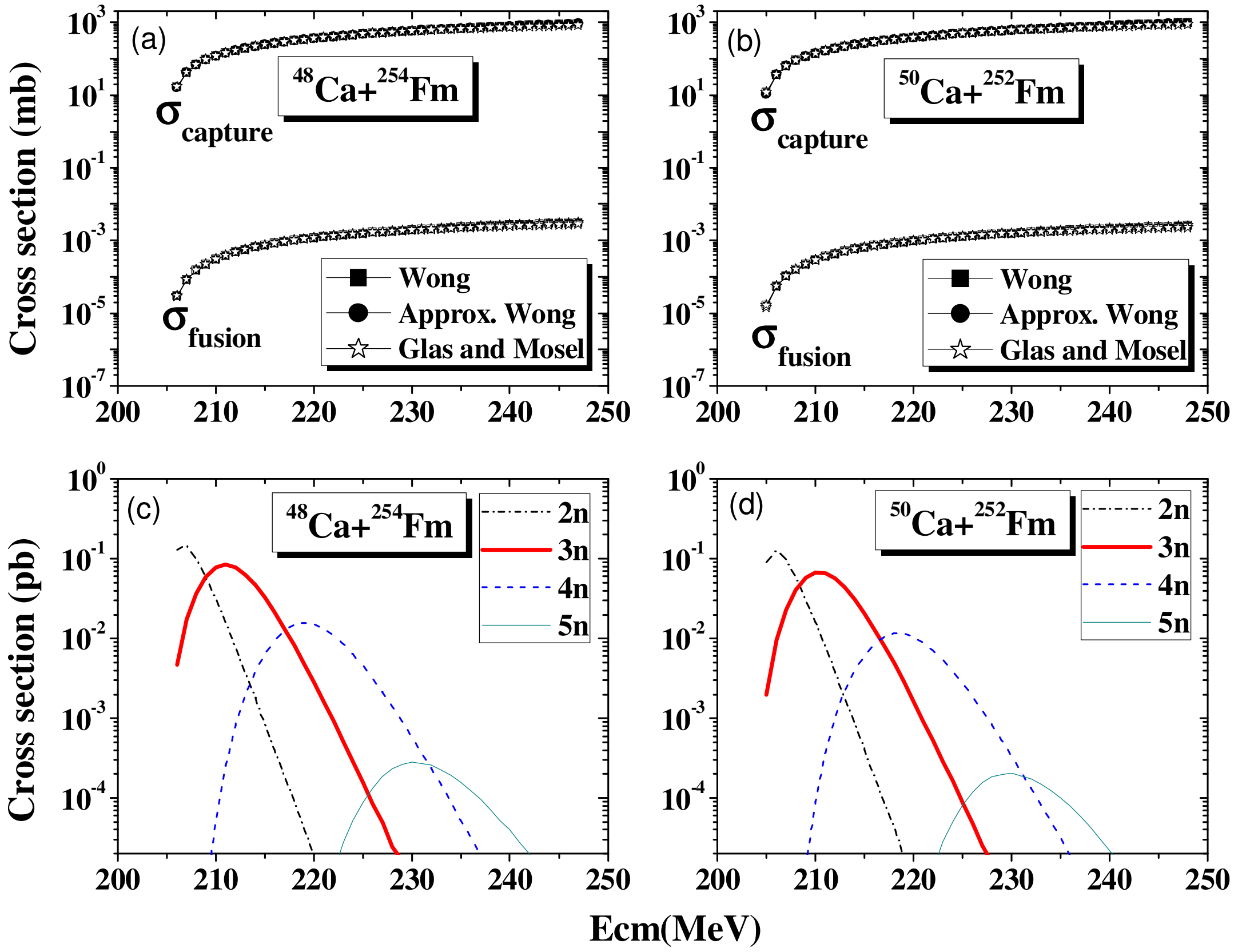

FIG 15. Plots of capture ($\sigma_{capture}$) and fusion ($\sigma_{fusion}$) cross sections in *mb* (upper panel) and evaporation residue cross section in *pb* (lower panel) vs. center of mass energy *Ecm* in MeV for the reactions of $^{48}$Ca+$^{254}$Fm and $^{50}$Ca+$^{252}$Fm systems

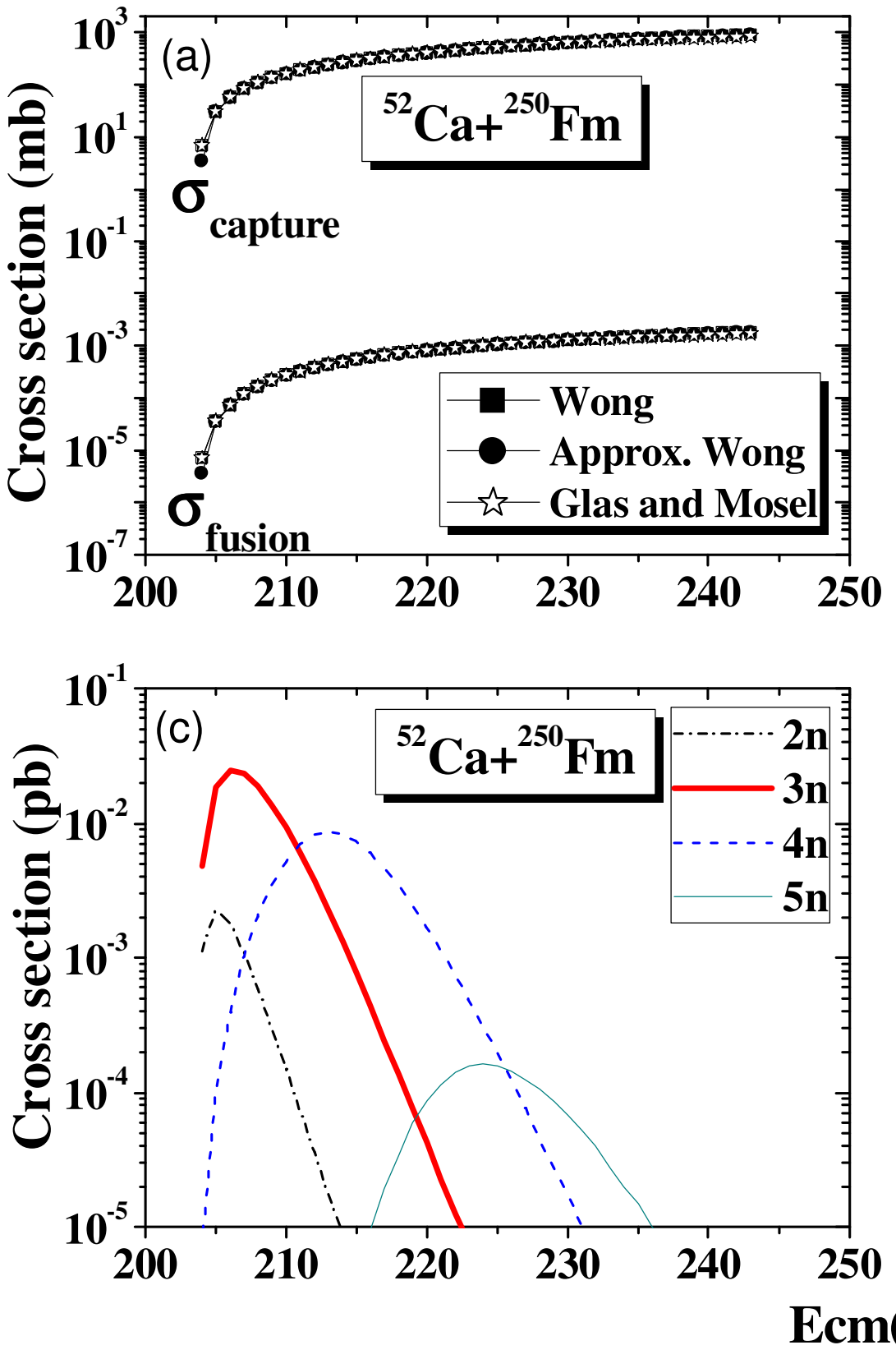

FIG 16.Plots of capture ($\sigma_{capture}$) and fusion ($\sigma_{fusion}$) cross sections in *mb* (upper panel) and evaporation residue cross section in *pb* (lower panel) vs. center of mass energy *Ecm* in MeV for the reactions of $^{52}$Ca + $^{250}$Fm and $^{54}$Ti + $^{248}$Cf systems.

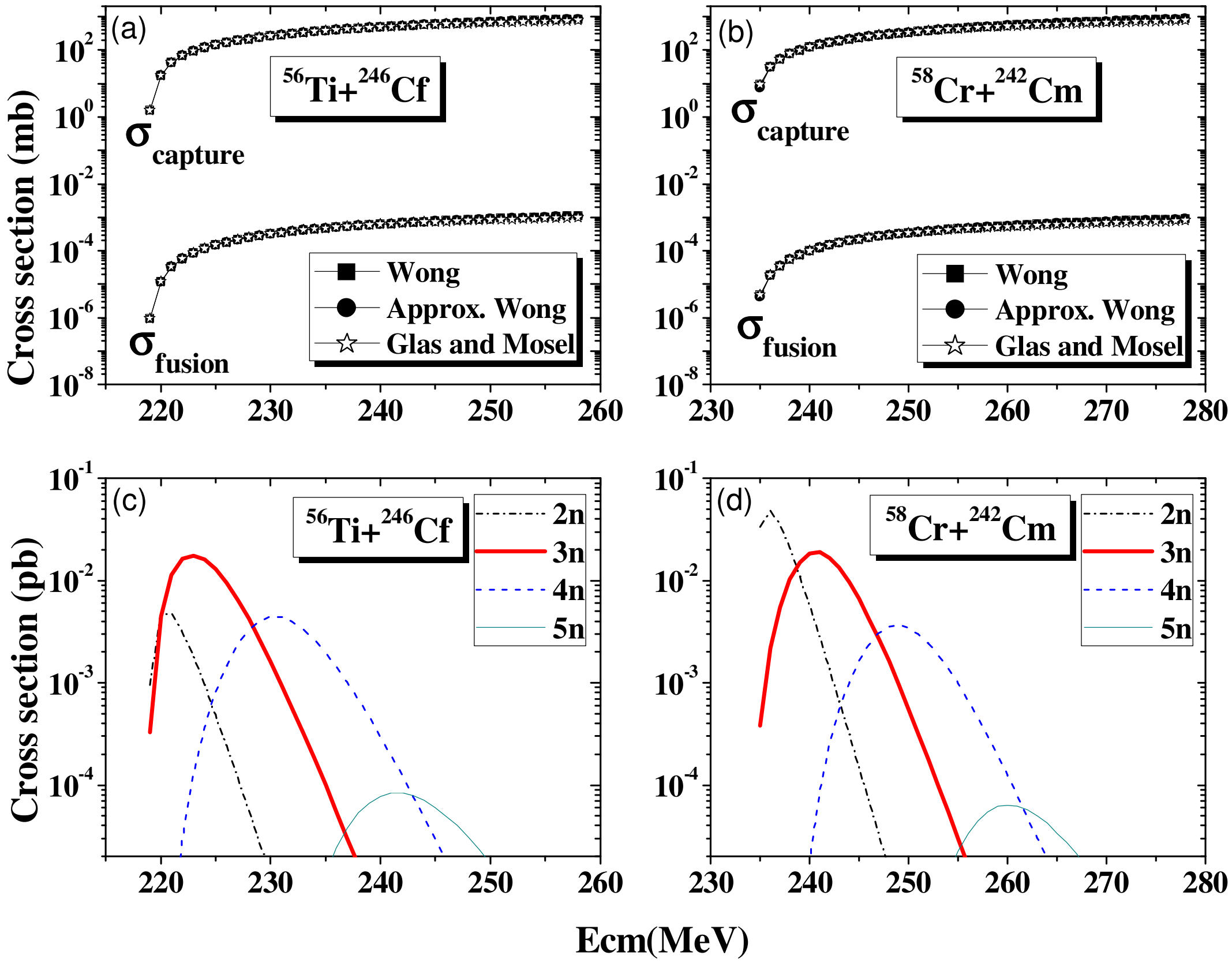

FIG 17. Plots of capture ($\sigma_{capture}$) and fusion ($\sigma_{fusion}$) cross sections in *mb* (upper panel) and evaporation residue cross section in *pb* (lower panel) vs. center of mass energy *Ecm* in MeV for the reactions of $^{56}$Ti+$^{246}$Cf and $^{58}$Cr+$^{244}$Cm systems.

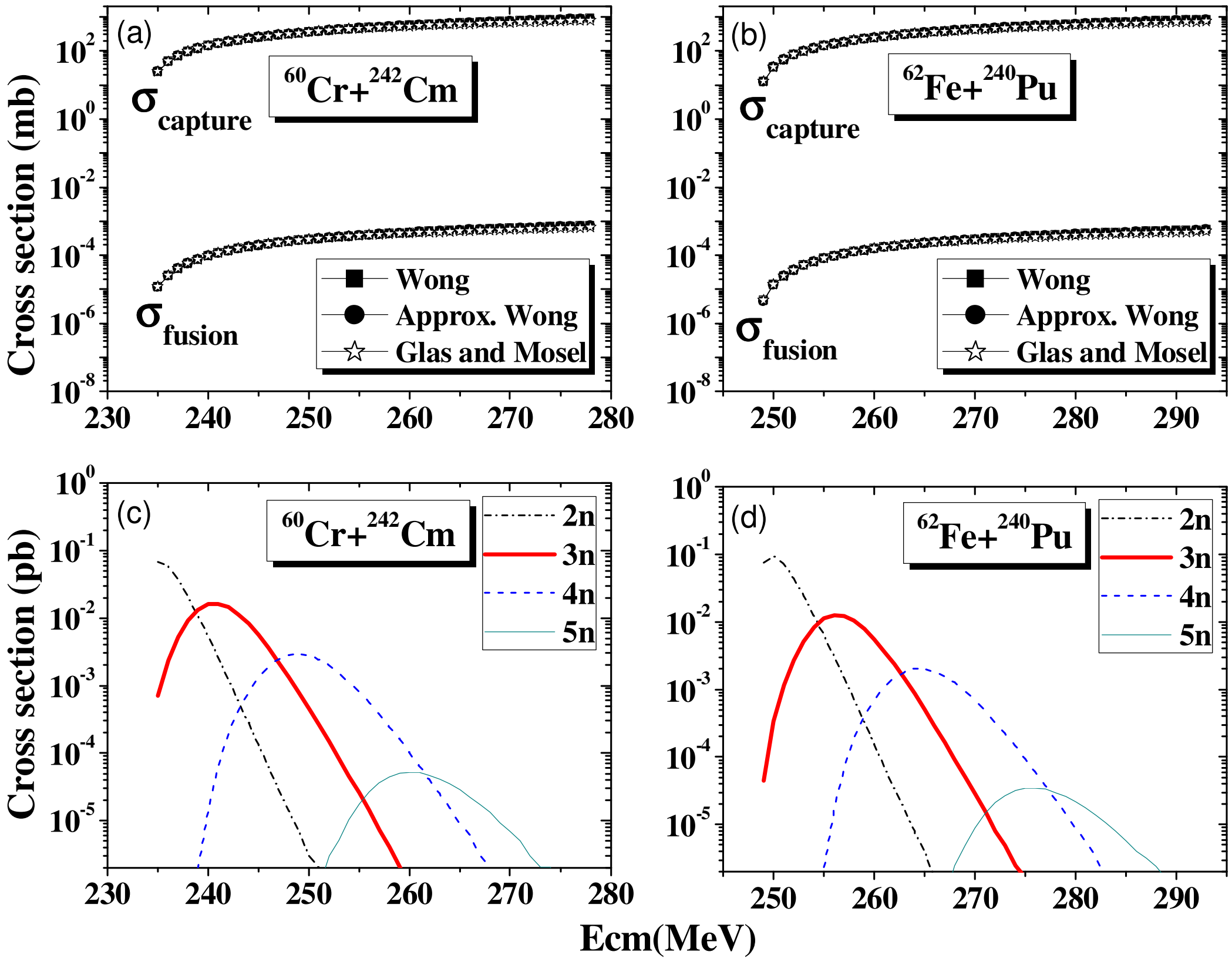

FIG 18. Plots of capture ($\sigma_{capture}$) and fusion ($\sigma_{fusion}$) cross sections in *mb* (upper panel) and evaporation residue cross section in *pb* (lower panel) vs. center of mass energy *Ecm* in MeV for the reactions of $^{60}$Cr + $^{242}$Cm and $^{62}$Fe + $^{240}$Pu systems.

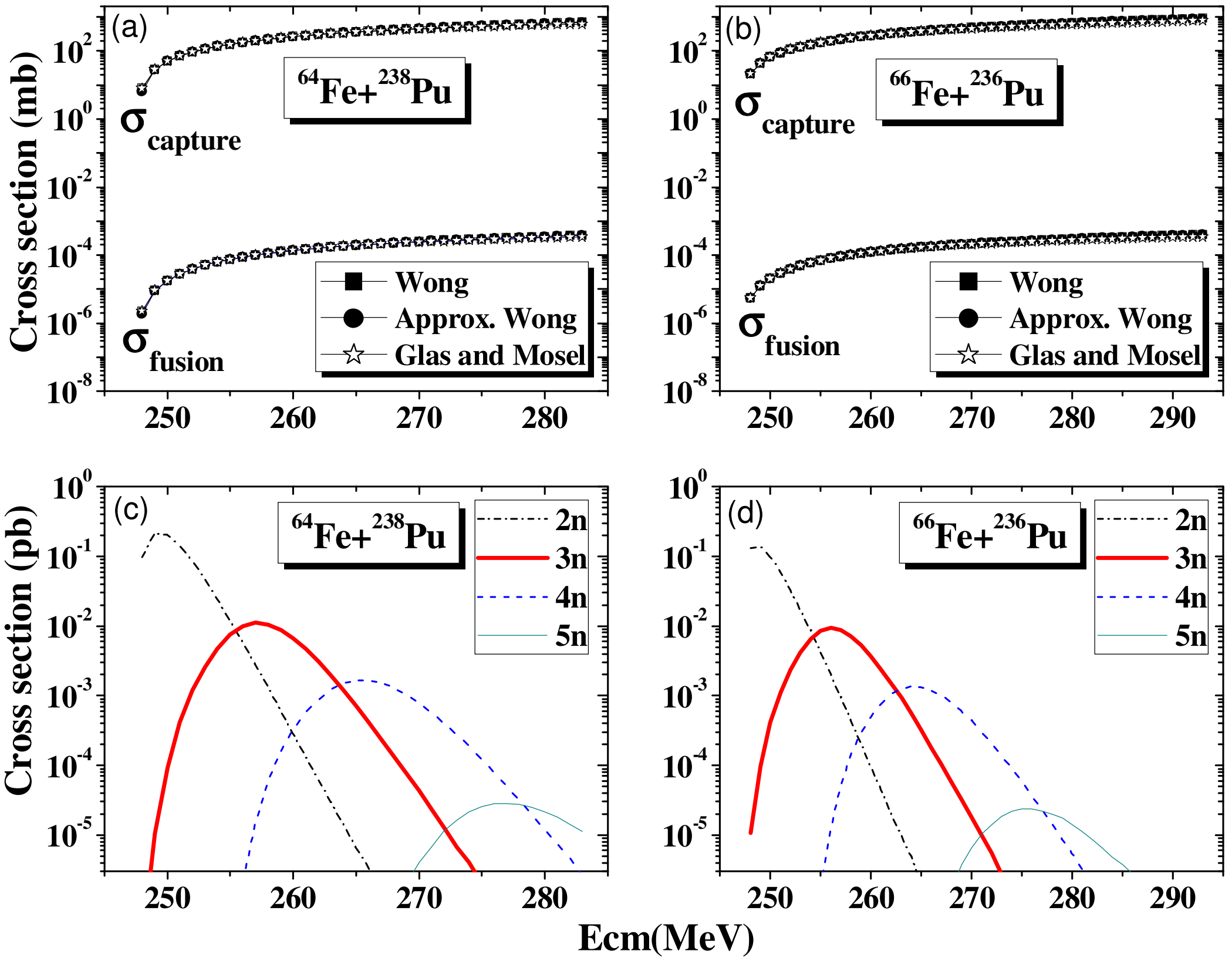

FIG 19.Plots of capture ($\sigma_{capture}$) and fusion ($\sigma_{fusion}$) cross sections in *mb* (upper panel) and evaporation residue cross section in *pb* (lower panel) vs. center of mass energy *Ecm* in MeV for the reactions of $^{64}$Fe + $^{238}$Pu and $^{66}$Fe + $^{236}$Pu systems.

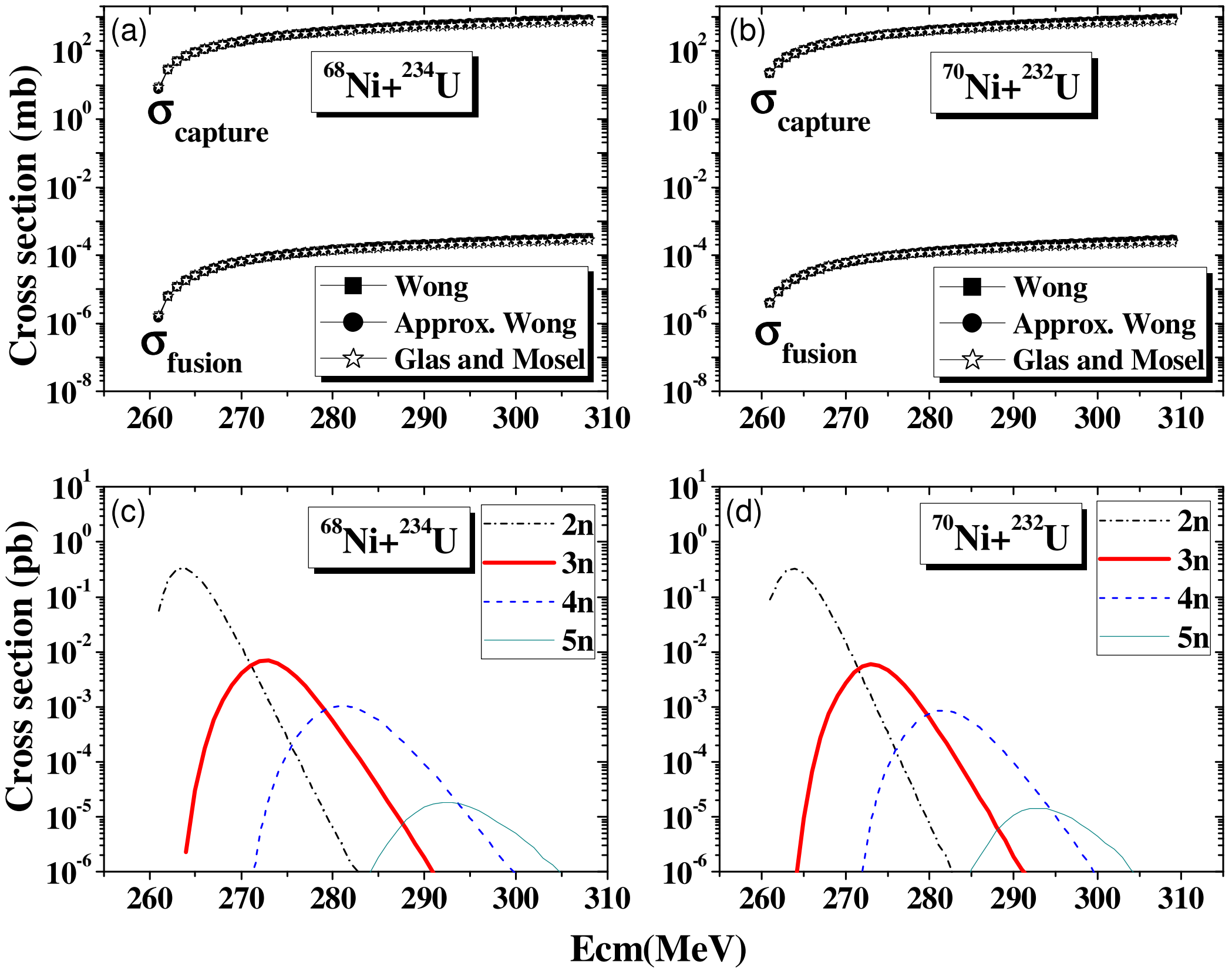

FIG 20. Plots of capture ($\sigma_{capture}$) and fusion ($\sigma_{fusion}$) cross sections in *mb* (upper panel) and evaporation residue cross section in *pb* (lower panel) vs. center of mass energy *Ecm* in MeV for the reactions of $^{68}$Ni + $^{234}$U and $^{70}$Ni + $^{232}$U systems.

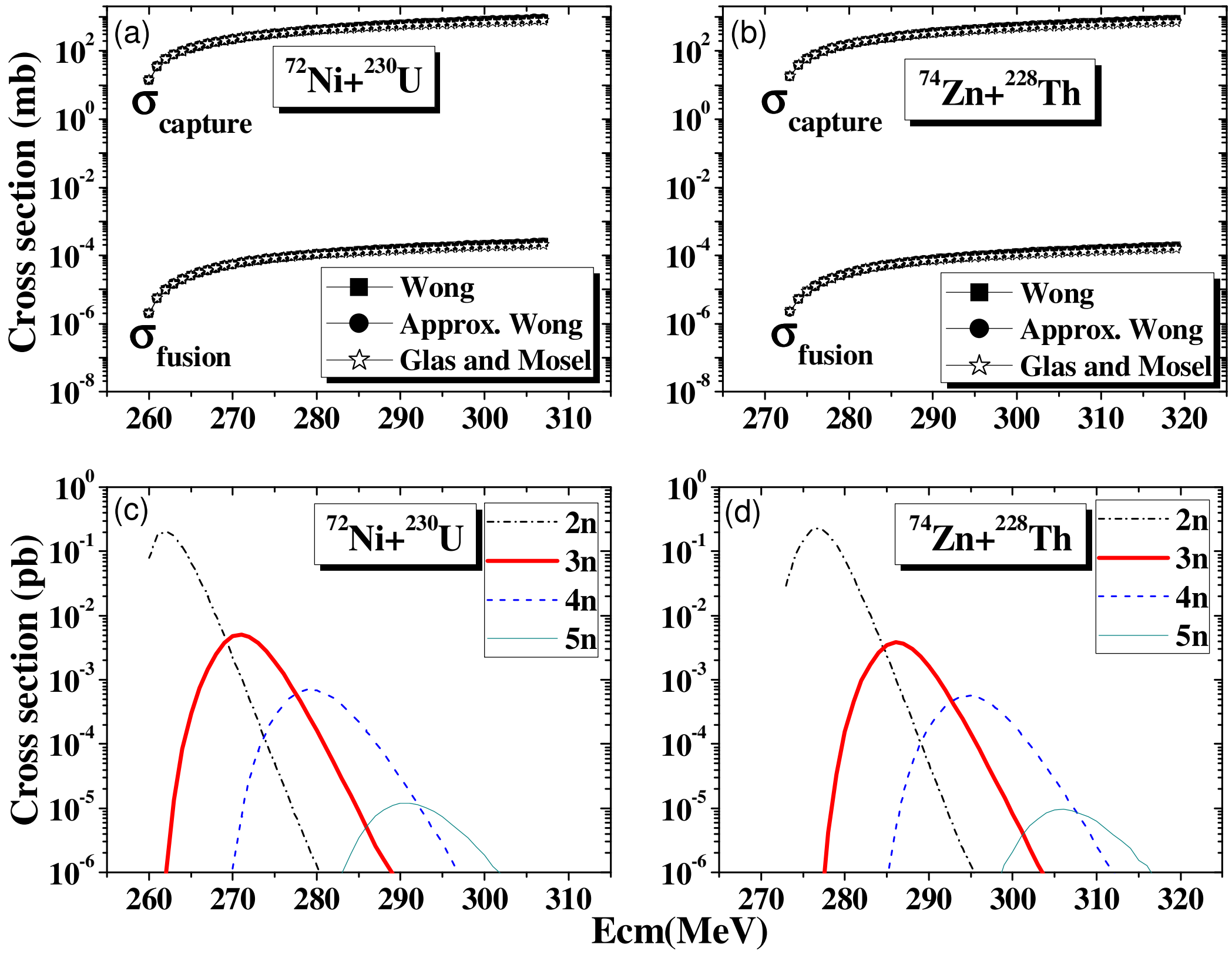

FIG 21. Plots of capture ($\sigma_{capture}$) and fusion ($\sigma_{fusion}$) cross sections in *mb* (upper panel) and evaporation residue cross section in *pb* (lower panel) vs. center of mass energy *Ecm* in MeV for the reactions of $^{72}$Ni+$^{230}$U and $^{74}$Zn+$^{228}$ Th systems.

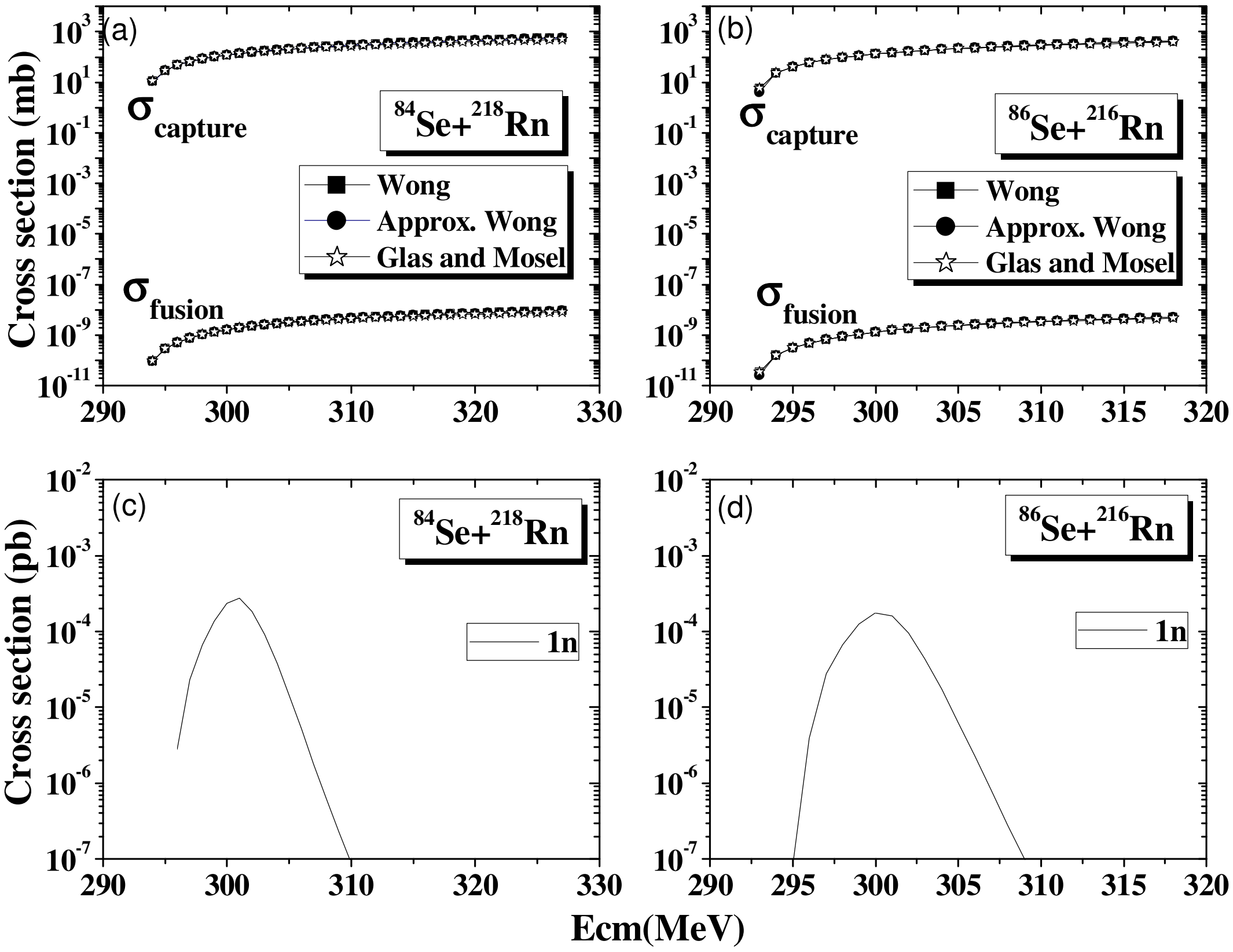

FIG 22. Plots of capture ($\sigma_{capture}$) and fusion ($\sigma_{fusion}$) cross sections in *mb* (upper panel) and evaporation residue cross section in *pb*(lower panel) vs. center of mass energy *Ecm* in MeV for the reactions of $^{84}$Se + $^{218}$Rn and $^{86}$Se +$^{216}$Rn of systems.

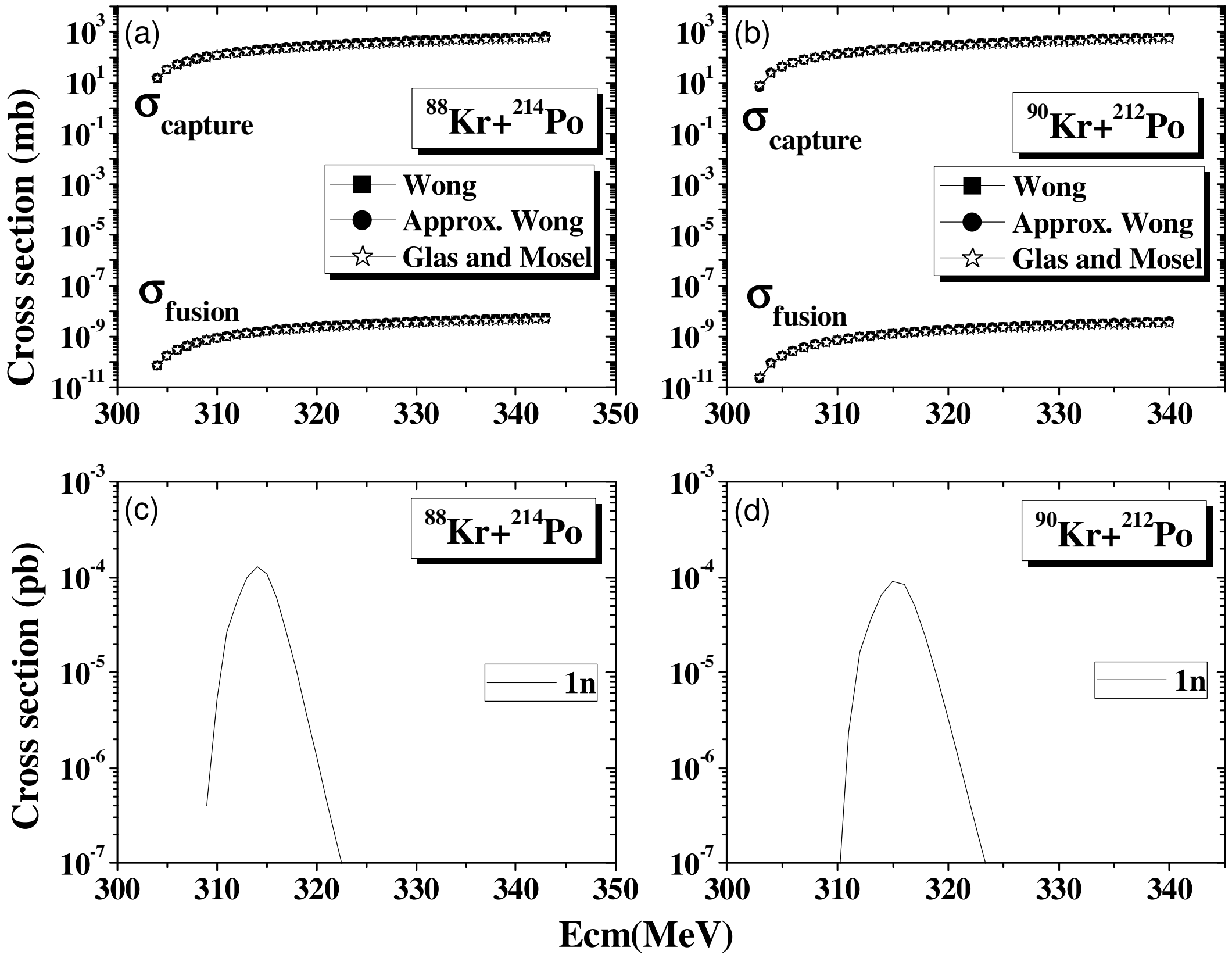

FIG 23. Plots of capture ($\sigma_{capture}$) and fusion ($\sigma_{fusion}$) cross sections in *mb* (upper panel) and evaporation residue cross section in *pb* (lower panel) vs. center of mass energy *Ecm* in MeV for the reactions of $^{88}$Kr + $^{214}$Po and $^{90}$Kr + $^{212}$Po of systems.

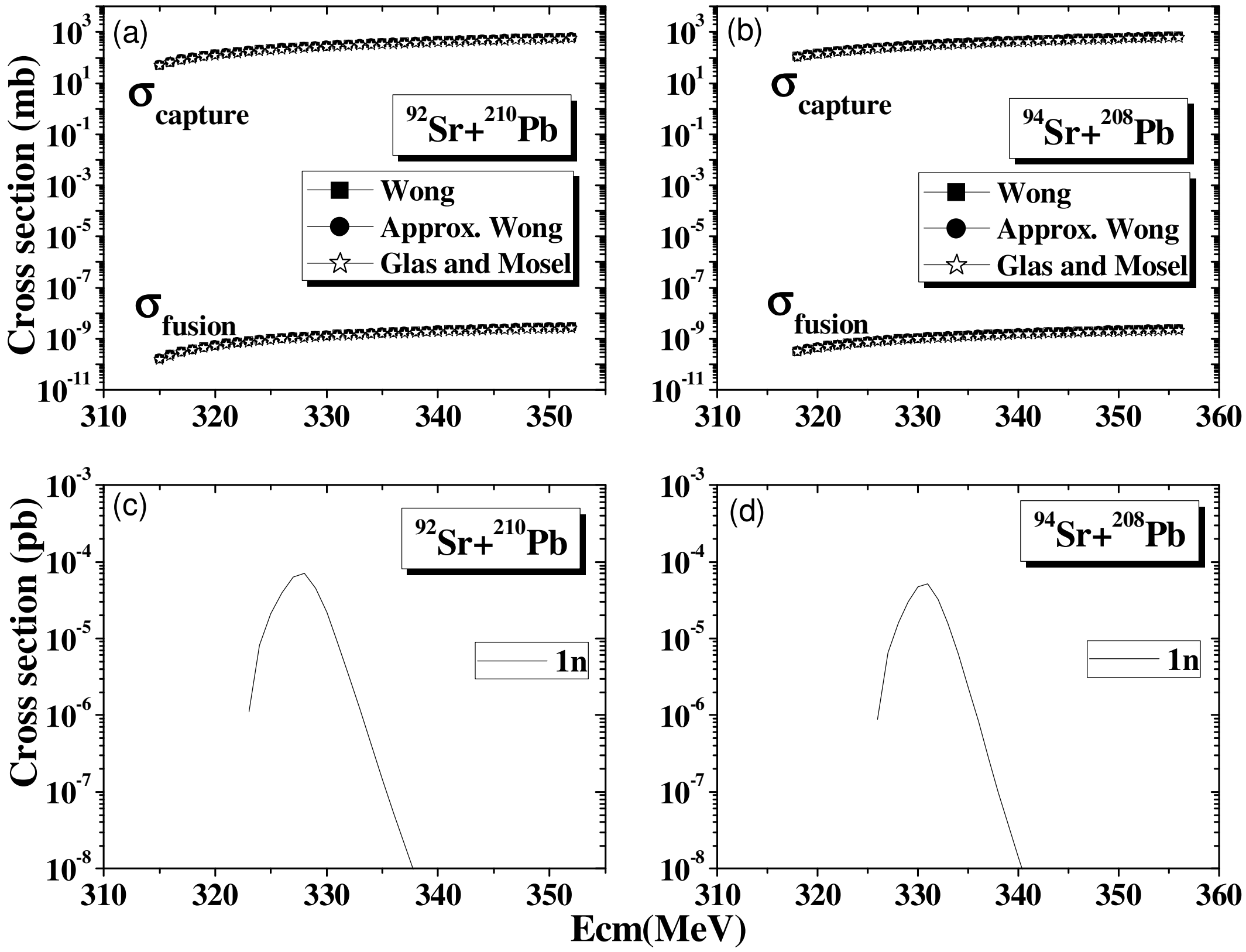

FIG 24.Plots of capture ($\sigma_{capture}$) and fusion ($\sigma_{fusion}$) cross sectionsin*mb* (upper panel) and evaporation residue cross section in *pb* (lower panel) vs. center of mass energy *Ecm* in MeV for the reactions of $^{92}$Sr + $^{210}$Pb and $^{94}$Sr + $^{208}$Pb systems.

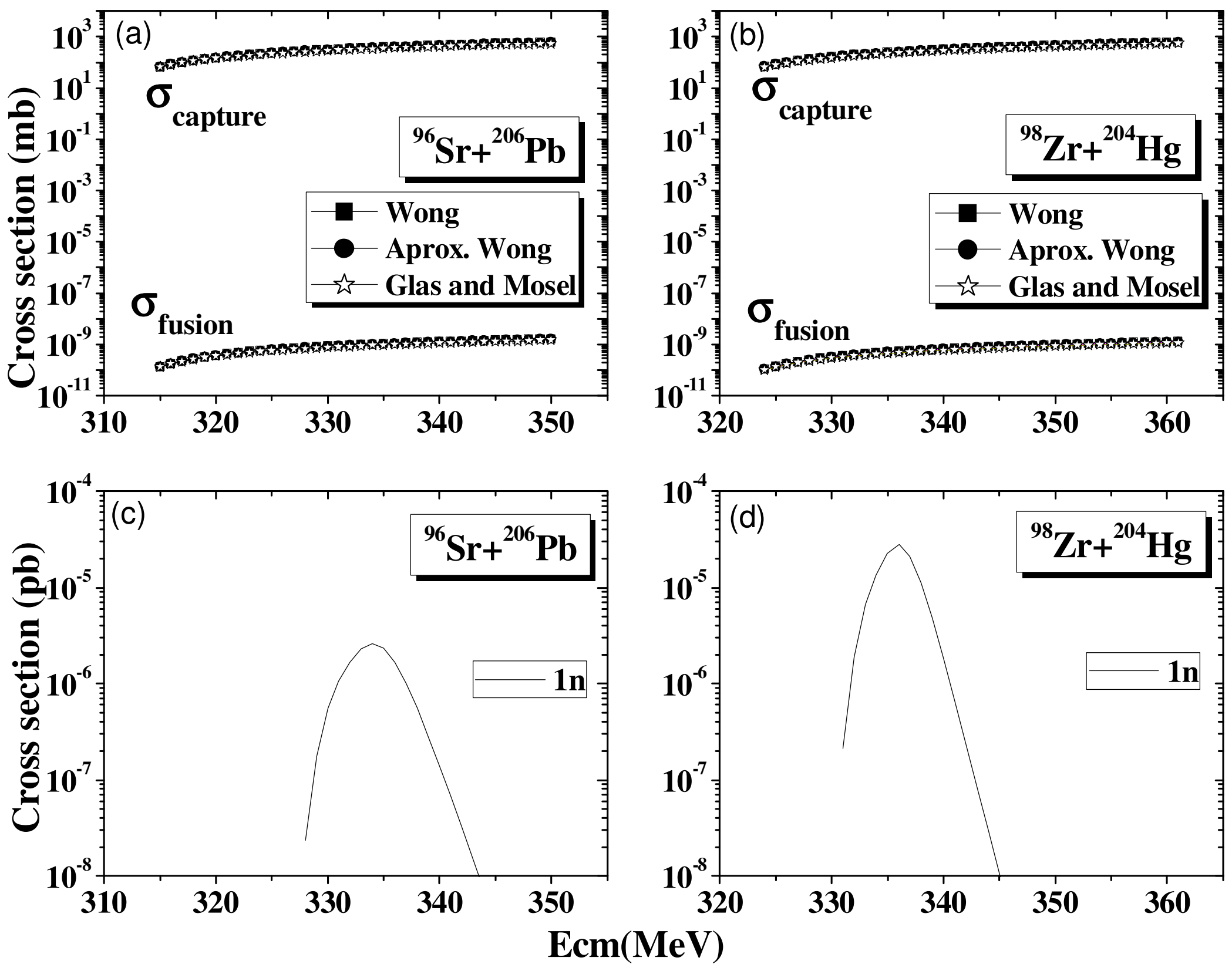

FIG 25.Plots of capture ($\sigma_{capture}$) and fusion ($\sigma_{fusion}$) cross sections in *mb* (upper panel) and evaporation residue cross section in *pb* (lower panel) vs. center of mass energy *Ecm* in MeV for the reactions of $^{96}$Sr + $^{206}$Pb and $^{98}$Zr + $^{204}$Hg systems.

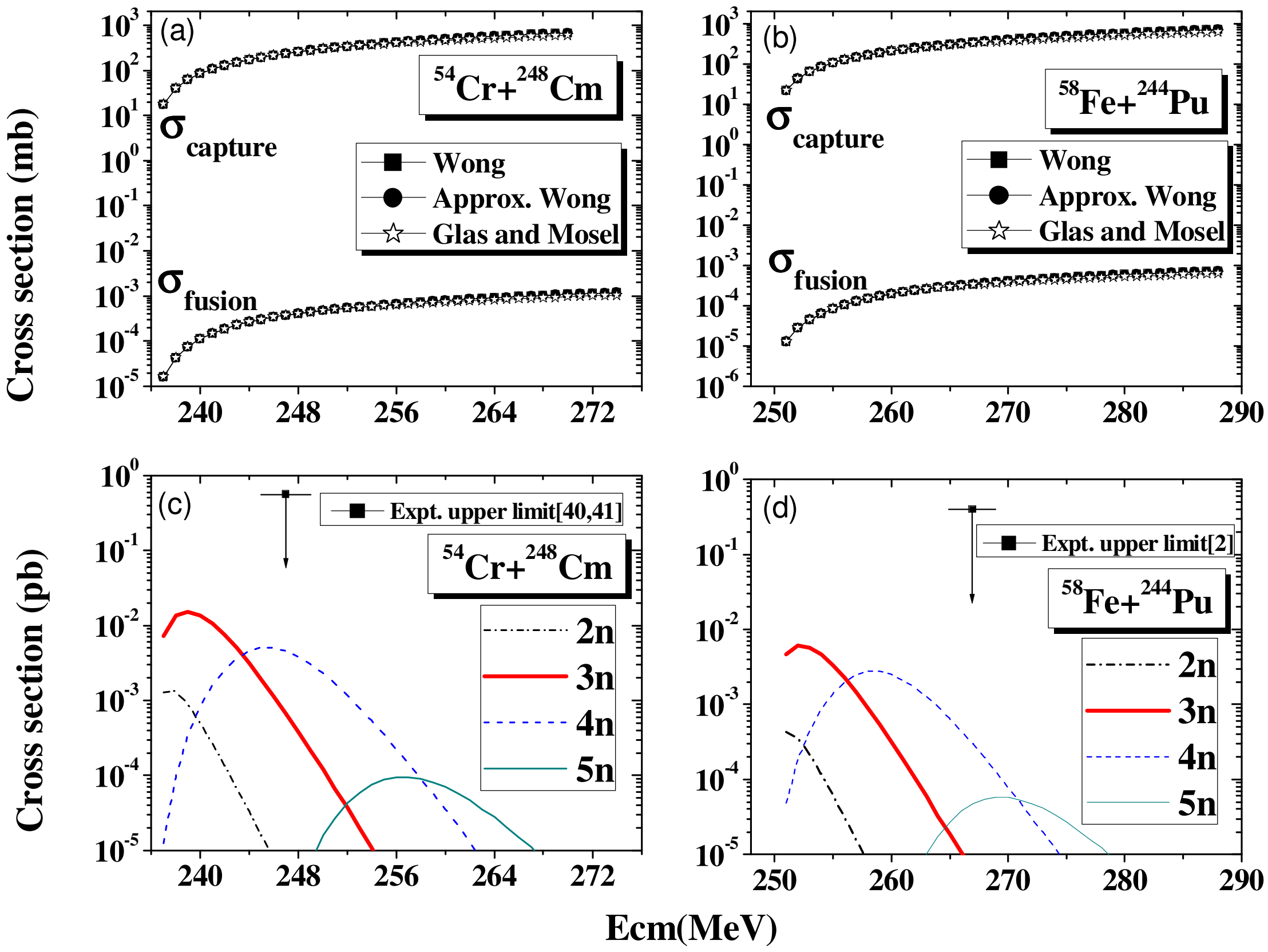

FIG 26.Plots of capture ($\sigma_{capture}$) and fusion ($\sigma_{fusion}$) cross sections in *mb* (upper panel) and evaporation residue cross section in *pb* (lower panel) vs. center of mass energy *Ecm* in MeV for the reactions of $^{54}$Cr + $^{248}$Cm and $^{58}$Fe + $^{244}$Pu systems.

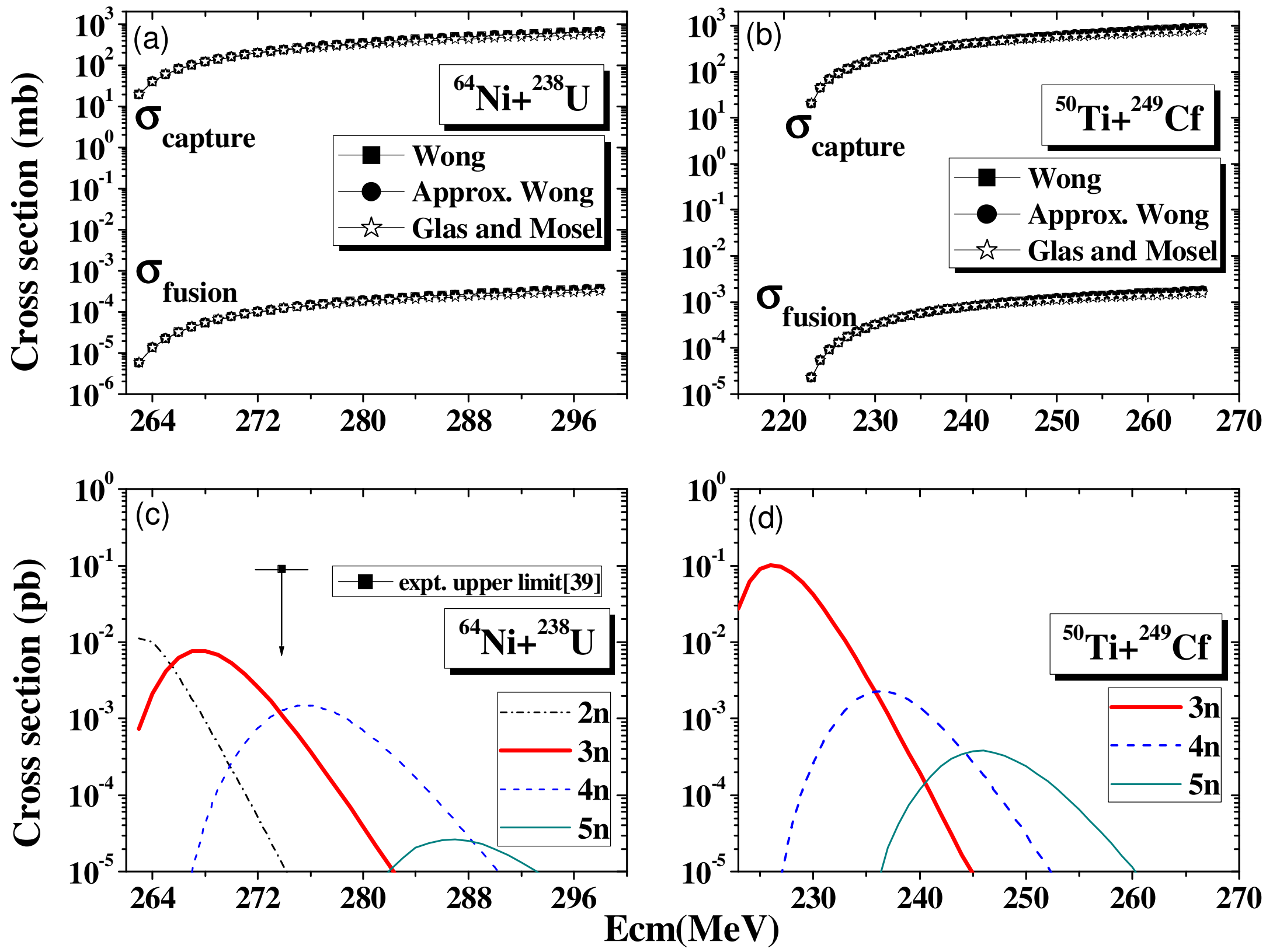

FIG 27. Plots of capture ($\sigma_{capture}$) and fusion ($\sigma_{fusion}$) cross sections in *mb* (upper panel) and evaporation residue cross section in *pb* (lower panel) vs. center of mass energy *Ecm* in MeV for the reactions of $^{64}$Cr + $^{248}$Cm and $^{58}$Fe + $^{244}$Pu systems.

TABLE 2. Calculated maximum values of the evaporation residues cross section for the probable hot and cold fusion combinations found in the cold valley of $^{302}120$.

| Combinations for hot fusion reactions | $\sigma_{ER}(2n)$ (fb) | $\sigma_{ER}(3n)$ (fb) | $\sigma_{ER}(4n)$ (fb) | $\sigma_{ER}(5n)$ (fb) | Combinations for cold fusion reactions | $\sigma_{ER}(1n)$ (fb) |
|---|---|---|---|---|---|---|
| $^{44}Ar+^{258}No$ | 2.3513 | 54.210 | 27.393 | 0.5343 | $^{84}Se+^{218}Rn$ | 0.2740 |
| $^{46}Ar+^{256}No$ | 7.4563 | 72.232 | 22.063 | 0.4020 | $^{86}Se+^{216}Rn$ | 0.1753 |
| $^{48}Ca+^{254}Fm$ | 146.89 | 84.207 | 15.669 | 0.2776 | $^{88}Kr+^{214}Po$ | 0.1283 |
| $^{50}Ca+^{252}Fm$ | 130.83 | 67.579 | 11.536 | 0.2020 | $^{90}Kr+^{212}Po$ | 0.0906 |
| $^{52}Ca+^{250}Fm$ | 2.2783 | 24.631 | 8.7108 | 0.1629 | $^{92}Sr+^{210}Pb$ | 0.0715 |
| $^{54}Ti+^{248}Cf$ | 78.024 | 34.790 | 6.4196 | 0.1116 | $^{94}Sr+^{208}Pb$ | 0.0523 |
| $^{56}Ti+^{246}Cf$ | 4.7816 | 17.639 | 4.4012 | 0.0844 | $^{96}Sr+^{206}Pb$ | 0.0406 |
| $^{58}Cr+^{244}Cm$ | 48.458 | 18.865 | 3.6606 | 0.0638 | $^{98}Zr+^{204}Hg$ | 0.0281 |
| $^{60}Cr+^{242}Cm$ | 67.820 | 16.252 | 2.9514 | 0.0521 | | |
| $^{62}Fe+^{240}Pu$ | 96.964 | 12.805 | 2.0295 | 0.0350 | | |
| $^{64}Fe+^{238}Pu$ | 215.07 | 11.067 | 1.6679 | 0.0282 | | |
| $^{66}Fe+^{236}Pu$ | 139.08 | 9.2581 | 1.3610 | 0.0233 | | |
| $^{68}Ni+^{234}U$ | 325.21 | 6.9364 | 1.0514 | 0.0183 | | |
| $^{70}Ni+^{232}U$ | 303.75 | 6.0200 | 0.8626 | 0.0144 | | |
| $^{72}Ni+^{230}U$ | 209.49 | 5.1034 | 0.7114 | 0.0121 | | |
| $^{74}Zn+^{228}Th$ | 228.78 | 3.8634 | 0.5699 | 0.0097 | | |
| $^{54}Cr+^{248}Cm$* | 1.3288 | 15.398 | 5.0357 | 0.0950 | | |
| $^{58}Fe+^{244}Pu$* | 0.4329 | 6.1166 | 2.7777 | 0.0580 | | |
| $^{64}Ni+^{238}U$* | 11.250 | 7.8063 | 1.4923 | 0.0267 | | |
| $^{50}Ti+^{249}Cf$* | 0.4140 | 102.92 | 2.2642 | 0.3784 | | |

*Combinations for which experimental studies were done.

TABLE 3.Comparison of predicted maximum values of the evaporation residues cross section (*$\sigma_{ER}$(3n)* and *$\sigma_{ER}$(4n)*) for $^{54}$Cr+$^{248}$Cm, $^{58}$Fe+$^{244}$Pu with our results. The maximum values of the presented data were taken from the figures of the ER excitation functions of the given references.

| $^{54}$Cr+$^{248}$Cm | | | | $^{58}$Fe+$^{244}$Pu | | | | Reference |
|---|---|---|---|---|---|---|---|---|
| Ecm (MeV) | $\sigma_{ER}$(3n) (fb) | Ecm (MeV) | $\sigma_{ER}$(4n) (fb) | Ecm (MeV) | $\sigma_{ER}$(3n) (fb) | Ecm (MeV) | $\sigma_{ER}$(4n) (fb) | |
| 246.0 | 14.0 | 250.0 | 28.0 | 264.0 | 2.0 | 265.0 | 5.0 | [44] |
| 240.5 | 0.7 | 249.0 | 0.4 | - | - | - | - | [52] |
| 237.2 | 55.0 | 241.0 | 13.0 | - | - | - | - | [49] |
| 241.0 | 160 | 252.0 | 12.0 | 255.0 | 12.0 | 265.0 | 1.8 | [34] |
| 238.0 | 2.0 | 250.0 | 5.0 | 252.0 | 0.9 | 263.0 | 2.2 | [50] |
| 240.0 | 0.6 | 250.0 | 3.0 | - | - | - | - | [54] |
| 239.0 | 15.4 | 245.0 | 5.04 | 252.0 | 6.12 | 258.0 | 2.78 | This Work |

TABLE 4. Comparison of predicted maximum values of the evaporation residues cross section ($\sigma_{ER}(3n)$ and $\sigma_{ER}(4n)$) for $^{64}Ni+^{238}U$, $^{50}Ti+^{249}Cf$ with our results. The maximum values of the presented data were taken from the figures of the ER excitation functions of the given references.

| $^{64}Ni+^{238}U$ | | | | $^{50}Ti+^{249}Cf$ | | | | Reference |
|---|---|---|---|---|---|---|---|---|
| Ecm (MeV) | $\sigma_{ER}(3n)$ (fb) | Ecm (MeV) | $\sigma_{ER}(4n)$ (fb) | Ecm (MeV) | $\sigma_{ER}(3n)$ (fb) | Ecm (MeV) | $\sigma_{ER}(4n)$ (fb) | |
| 273.0 | 4.5 | 277.0 | 3.0 | 236.0 | 40 | 241 | 46 | [44] |
| - | - | - | - | 240.5 | 5.5 | 249 | 6.1 | [52] |
| - | - | - | - | 225.0 | 100 | 231.5 | 2.5 | [49] |
| 273.0 | 7.0 | 283.0 | 1.0 | 227.5 | 760 | 239 | 28 | [34] |
| - | - | - | - | 227.0 | 20 | 240 | 4.5 | [50] |
| - | - | - | - | 229.0 | 20 | 240 | 21 | [54] |
| - | - | - | - | 231.5 | 60 | 232.5 | 40 | [80] |
| - | - | - | - | 230.0 | 150 | 239 | 50 | [53] |
| - | - | - | - | 230.0 | 50 | 245 | 3.5 | [51] |
| 268.0 | 7.81 | 276.0 | 1.49 | 226.0 | 102.92 | 237 | 2.264 | This Work |